\documentclass[12pt]{iopart}
\usepackage{amsmath, amsfonts, amssymb, amsthm, amstext, bm, bbm, multirow, url, comment}
\usepackage{lmodern}
\usepackage[T1]{fontenc}

\usepackage{graphicx, subcaption, ragged2e}
\DeclareCaptionJustification{justified}{\justifying}
\usepackage[usenames, dvipsnames]{xcolor}
\usepackage[colorlinks=true, linkcolor=blue,citecolor=red, urlcolor=blue, hypertexnames=true]{hyperref}
\usepackage[textwidth=160mm, textheight = 210mm, hmarginratio={1:1}, vmarginratio={1:1}, 
			 heightrounded]{geometry} 
    
\newcommand{\brac}[1]{\left( #1 \right) }
\newcommand{\mathsfit}[1]{\textsf{\textit{#1}}}

\newcommand{\pdt}[0]{\partial_{t}}
\newcommand{\pd}[1]{\partial_{ #1 }}

\newcommand{\figref}[1]{Fig.~\ref{#1}}
\newcommand{\equref}[1]{Eq.~\eqref{#1}}

\renewcommand{\theequation}{\thesection.\arabic{equation}}

\begin{document}

\title{Universality in coupled stochastic Burgers systems with degenerate flux Jacobian}
 
\author{Dipankar Roy}
\address{International Centre for Theoretical Sciences, Tata Institute of Fundamental Research, Bangalore 560089, India}
\author{Abhishek Dhar}
\address{International Centre for Theoretical Sciences, Tata Institute of Fundamental Research, Bangalore 560089, India}
\author{Konstantin Khanin}
\address{Department of Mathematics, University of Toronto, 40 St George Street, Toronto, ON M5S 2E4, Canada}
\author{Manas Kulkarni}
\address{International Centre for Theoretical Sciences, Tata Institute of Fundamental Research, Bangalore 560089, India}
\author{Herbert Spohn}
\address{Zentrum Mathematik and Physik Department, Technische Universität München, Garching 85748, Germany}
\vspace*{22pt}
\noindent
\textbf{Abstract}. In our contribution we study stochastic models in one space dimension with two conservation laws. One model is the coupled continuum stochastic Burgers equation,
for which each current is a sum of quadratic non-linearities, linear diffusion, and spacetime white noise. The second model is a two-lane stochastic lattice gas. As distinct from previous
studies, the two conserved densities are tuned such that the flux Jacobian, a $2 \times 2$ matrix, 
has coinciding eigenvalues.
In the steady state, investigated are spacetime correlations of the conserved fields and the time-integrated currents at the origin. For a particular choice of couplings  the dynamical exponent 3/2 is  confirmed. Furthermore, at these couplings,  continuum stochastic Burgers equation  and lattice gas are demonstrated to be in the same universality class. 


\maketitle
\tableofcontents

\section{Introduction}
\label{sec1}
In the context of the one-dimensional Kardar-Parisi-Zhang (KPZ) equation governing surface growth \cite{1986-kardar--zhang, 1995-healy-zhang, 2012-corwin},
there is a long standing interest in stochastic models with a single conservation law.
A protoypical example is the continuum stochastic Burgers equation which describes the motion of KPZ slopes. Further examples are lattice gases, such as ASEP \cite{1998-derrida}, TASEP \cite{1993-derrida--pasquier, 2003-popkov-schutz, 2007-blythe-evans}, PNG \cite{2002-prahofer-spohn}, and more.
The extension to several components was largely triggered by the observation that 
the linear Landau-Lifshitz fluctuation theory fails for one-dimensional fluids. These classical particle systems generically have three conservation laws and the appropriate nonlinear version of the fluctuation theory 
is a continuum stochastic Burgers equation with three components~\cite{2014-spohn}. The analysis of these equations relies on the fact that the heat peak and sound peaks have distinct velocities.
More recently, as a great surprise, signatures of KPZ physics were also observed 
in the spin-spin spacetime correlations of the isotropic spin-$\tfrac{1}{2}$ chain
at high temperatures \cite{2011-znidaric, 2019-ljubotina--prosen, 2021-scheie--tennant, 2022-wei--zeiher}. In \cite{2023-nardis--vasseur} an effective theory for spin and helicity has been derived.
The dynamical equations are two coupled continuum Burgers equations. As a very particular feature, in this case
there are no sound modes and the standing still heat peak has two components, which leads to more intricate correlation effects and is not covered by the conventional analysis of  nonlinear fluctuating hydrodynamics \cite{2014-spohn}.
For us, this is one of the motivation to study more systematically stochastic Burgers systems, both continuum and lattice,
with a degenerate flux Jacobian.

For a single conserved field many detailed studies are available. Progress up to 2015 is discussed in the Les Houches lectures \cite{2016-spohn}. Surprising properties have been uncovered. Their experimental realization is reviewed in an instructive
article by K. Takeuchi \cite{ 2018-takeuchi}. In this contribution, we will focus on coupled stochastic continuum Burgers equations and lattice gases in one dimension with two conserved densities, $u_1$ and $u_2$. To understand properties of their spacetime dependent structure function the flux Jacobian is a central quantity. To provide a definition, we denote by $j_1(u_1,u_2)$ and $j_2(u_1,u_2)$ the  currents, averaged in a spatially homogeneous steady state, in dependence on the densities $\vec{u} = (u_1,u_2)$. The flux Jacobian is then the $2\times 2$ matrix
\begin{equation}
    \label{1.1}
    A_{\alpha\beta}(\vec{u})=\partial_{u_\beta}j_\alpha(\vec{u}),
\end{equation}
$\alpha,\beta =1,2$. More explicitly, the flux Jacobian takes the form
\begin{equation}
    \label{eq:fj-explicit}
    A = 
    \begin{pmatrix}
        \partial_{u_1} j_1 & \partial_{u_2} j_1 \\
        \partial_{u_1} j_2 & \partial_{u_2} j_2 
    \end{pmatrix}.
\end{equation}
This matrix depends on $\vec{u}$ and is  not symmetric, in general. However, $A$ has always real eigenvalues and a complete system of left and right  eigenvectors \cite{2011-grisi-schutz, 2013-ferrari--spohn, 2014-spohn}. 
If the lattice gas is prepared in a statistically stationary spatially homogeneous state, then an initial perturbation at the origin will generate two localized peaks travelling with velocities given by the eigenvalues of $A$. If $A$ is non-degenerate, for long times, the peaks are spatially far apart and one argues that the two propagating modes reduce to the case of a single component \cite{2013-mendl-spohn, 2014-das--spohn, 2014-spohn}. The fine structure of these peaks depends on the second order expansion of the currents and is handled by nonlinear fluctuating hydrodynamics. For generic  values of $\vec{u}$ the matrix $A$ will be non-degenerate. But there could be special points at which $A$ is degenerate. Then the peaks travel with the same velocity and the decoupling approximation fails. The understanding of this case  is the primary agenda of our investigation. 

An early contribution is the study by Erta\c{s} and Kardar \cite{1993-ertas-kardar}, see also \cite{1992-ertas-kardar}, who consider the dynamics of a stretched string, e.g. a vortex line, in a spacetime random medium. For the two transversal components they write two coupled stochastic differential equations of KPZ type and argue on physical grounds  that the flux Jacobian \eqref{eq:fj-explicit} vanishes. Erta\c{s} and Kardar report on numerical findings on scaling exponents for their coupled stochastic equations \cite{1992-ertas-kardar}. Given the fact that our understanding of one-component systems has significantly advanced since then \cite{2012-corwin, 2015-quastel-spohn, 2018-takeuchi}, it is worthwhile to reconsider the two-component case with degenerate flux Jacobian, in particular to move beyond scaling exponents. 

More recently revived interest has been initiated through the study of spin chains, both classical \cite{2013-prosen-zunkovic, 2019-das--dhar, 2020-krajnik-prosen, 2020-krajnik--prosen} and quantum \cite{2011-znidaric, 2019-ljubotina--prosen, 2022-ye--yao}. The KPZ exponent $3/2$ was theoretically established and experimentally confirmed \cite{2021-scheie--tennant,2022-wei--zeiher,2023-google-qu-ai}.
Suprisingly, theoretical studies~\cite{2019-ljubotina--prosen,2019-das--dhar,2020-krajnik-prosen, 2020-krajnik--prosen} have also found that the spacetime spin-spin correlator shows a scaling function remarkably close 
to the exact solution from the one-component stationary KPZ equation.
In an attempt to understand such features a coarse-grained description for the dynamics of spin and helicity density has been derived in \cite{2023-nardis--vasseur}. 
The resulting model is stochastic and has  the form of a two-component stochastic Burgers equation with degenerate flux Jacobian.
 

In our contribution, we report on numerical studies of  continuum stochastic Burgers equations with two components, in which case
degeneracy  of the flux Jacobian is easily implemented. 
We also investigate  a stochastic lattice gas with two lanes \cite{2014-popkov--schutz} and notice a remarkable universality between lattice and continuum systems.
In the lattice gas, the particles stay in their own lane, hence the lane densities are conserved. The interaction arises  through jump rates depending on the respective occupations in the other lane.  At densities $u_1 = \tfrac{1}{2} = u_2$, the flux Jacobian turns out to be degenerate. In the  continuum and lattice models the spatially homogeneous stationary state is known explicitly. We then investigate the spacetime covariance, a $2\times 2$ matrix, of the conserved fields in the stationary state. Studied is also the time-integrated current through the origin, which  is a random two-vector, up to  a term linear in time. For a single component, the scaling form of both observables is known rigorously and our task is to elucidate  their extension to two components.

The paper is organized as follows. In Section~\ref{sec2},  we discuss continuum coupled stochastic Burgers equations and study the cyclicity condition, ensuring a time-stationary measure which is Gaussian.  The two-lane lattice model is introduced in Section~\ref{sec4} and its intriguing connection to coupled Burgers equations is established.
Our numerical results on a discretized version of the two-component stochastic Burgers equation are presented in Section~\ref{sec3}, while the Monte-Carlo results for the two-lane stochastic lattice gas model are provided in Section~\ref{sec5}. Universality is taken up in Section~\ref{sec6}. We also recall in Section~\ref{sec7} the mode-coupling theory for two-component stochastic Burgers equations. 
A more detailed comparison to prior work, discussion points,  and summary can be found towards the end of the article. Some computations are relegated to the appendices.

\section{Coupled stochastic Burgers equations, cyclicity condition}
\label{sec2}
\setcounter{equation}{0}

As starting point of our investigations we consider 
two coupled Langevin equations of stochastic Burgers type. The well-understood case of one component is briefly recalled.
To have a concrete set-up for two components,  
the focus in this section is primarily on time-stationary steady states and on linear transformations of the two random density fields.
\subsection{One component}
\label{sec2.1}
The stochastic Burgers equation with a single component reads 
\begin{equation}
    \label{2.1}
    \partial_t \phi(x,t) - \partial_x\big(\mathsfit{a} \phi(x,t) + \lambda\phi(x,t)^2 +  \mathsfit{d}\partial_x \phi(x,t) + \mathsfit{b} \,\xi(x,t) \big) = 0.
\end{equation}
Here $\phi(x,t)$ is the density field and $\xi(x,t)$ is standard spacetime Gaussian white noise,
\begin{equation}
    \label{2.2}
    \langle \xi(x,t) \xi(x',t')\rangle =   \delta( x - x' ) \delta(t-t').
\end{equation}
In analysing the Langevin equation \eqref{2.1}, a major simplification results from the knowledge of the time-stationary measures. In fact, these are Gaussian measures with arbitrary mean and covariance $\chi\delta(x-x')$. The susceptibility $\chi$ is determined by
\begin{equation}
    \label{2.3}
    \chi   =  \mathsfit{b}^2/ 2\mathsfit{d}. 
\end{equation}
To be noted,  $\chi$ does not depend on the strength of the quadratic nonlinearity. Shifting the field $\phi$ by a constant as $\phi +\kappa_0$ merely modifies $\mathsfit{a}$ as $\mathsfit{a} + 2\lambda\kappa_0$. In turn
the linear term can be removed by a Galilei transformation. Thus we can set $\mathsfit{a}=0$ and 
$\langle\phi(x)\rangle =0$, average in the spatially homogeneous steady state.

The observable of interest is the spacetime correlator
\begin{equation}
    \label{2.4}
    \mathsfit{S}(x,t) = \langle\phi(x,t)\phi(0,0)\rangle, 
\end{equation}
which, because of the conservation law, satisfies the sum rule
\begin{equation}
    \label{2.5}
    \int_\mathbb{R} dx \mathsfit{S}(x,t) = \chi.
\end{equation}
Asymptotically  $\mathsfit{S}(x,t)$ is of scaling form. To determine the relation between space and time, one rescales $\phi(x,t)$. As a result the dynamical exponent $z$ takes the value $z=3/2$. Diffusion and noise term are sub-leading, but ensure local time-stationarity. For the correlator these considerations  imply the scaling form
\begin{equation}
    \label{2.6}
    \mathsfit{S}(x,t) = \chi\, t^{-2/3}\mathsfit{g}(t^{-2/3}x).
\end{equation}
for large $x,t$. The integral of $\mathsfit{g}$ equals one and the prefactor ensures the validity of  \equref{2.5}.

As one important discovery from 2004 \cite{2004-prahofer-spohn},  the scaling function can be computed exactly with the result
\begin{equation}
    \label{2.7}
    \mathsfit{S}(x,t) \simeq \chi (\Gamma_\| t)^{-2/3}f_\mathrm{KPZ}\big((\Gamma_\| t)^{-2/3}x\big)
    \medskip
\end{equation}
with $\Gamma_\| = 2|\lambda| \sqrt{2\chi}$. A numerical plot of the function $f_\mathrm{KPZ}$ can be found in \cite{2004-prahofer-spohn}. 
The KPZ scaling theory claims that \equref{2.7} holds for a wide range of stochastic particle models with a single conservation law, where $\chi$ is the static susceptibility and $\lambda = j''(\rho)/2$ with $j(\rho)$
the steady state average current. In this sense $f_\mathrm{KPZ}$ is a universal function and $\Gamma_\|$ a very specific 
model-dependent parameter.
  In fact, the scaling \eqref{2.7} was originally obtained for the polynuclear growth model and later confirmed for the TASEP \cite{2006-ferrari-spohn}. For the continuum stochastic Burgers equation \eqref{2.1}, the scaling \eqref{2.7} was proved only eight years later \cite{2015-borodin--veto}. 

\subsection{Two components}
\label{sec2.2}
 The formal extension of \equref{2.1} to two components becomes
\begin{equation}
    \label{2.8}
    \partial_t \phi_\alpha - \partial_x\big(A_{\alpha\beta}\phi_\beta + \langle\vec{\phi},  G^\alpha \vec{\phi} \,\rangle+ D_{\alpha\beta}\partial_x\phi_\beta + B_{\alpha\beta}\xi_\beta\,\big) =0,
\end{equation}
$\alpha = 1,2$, where $\xi_1,\xi_2$ are independent standard white noise components and $G^1 , G^2$ are constant $2 \times 2$ matrices, which are symmetric by construction. 
Here and below, we use the Einstein convention for summing over repeated Greek indices,
i.e. summing over $\beta = 1,2$ above. Sometimes it is convenient to use vector notation as 
\begin{equation}
    \label{2.9}
    \vec{\phi} = (\phi_1, \phi_2),  \qquad
    \vec{\xi}       = (\xi_1,\xi_2),  \qquad  \vec{G}       = (G^1,G^2).
\end{equation}
In \equref{2.8}, the symbol $\langle\cdot,\cdot\rangle$ then refers the scalar product for $2$-vectors and $A,D,B,G^\alpha$ are constant $2 \times 2$ matrices. For the degenerate case of interest, $A = c_01$ and $c_0$ can be set to zero through a Galilei transformation.
Hence we set $A=0$.

As in the case of a single component, one would like to determine the time-stationary measure. Since the system 
is nonlinear and does not satisfy detailed balance, no generic tools are available. One way out is to study the reverse question. We first fix  a Gaussian measure with mean zero and covariance 
\begin{equation}
    \label{2.10}
    \langle \phi_\alpha(x) \phi_{\alpha'}(x')\rangle =   C_{\alpha\alpha'}\delta( x - x' ), 
\end{equation}
where $C$ is a symmetric strictly positive definite $2\times2$ matrix. As a generalization of \equref{2.3}, for the linear part of the coupled equation one has the relation
\begin{equation}
    \label{2.11}
    DC + CD^{\mathrm{T}}  = BB^{\mathrm{T}} 
\end{equation}
with $^{\mathrm{T}}$ denoting transpose. The Gaussian measure with covariance \eqref{2.10} is time-stationary in case $\vec{G} = 0$. Therefore we view $C$ as given, but still allowing for some freedom of how to choose the matrices $B,D$. However, the linear part should have good mixing properties, which is ensured by $BB^\mathrm{T} > 0$
and hence also $ D + D^\mathrm{T} > 0$. The issue is now to determine for which couplings $\vec{G}$ the Gaussian measure remains time-stationary. 
This question can be resolved and we first state the result. One defines
\begin{equation}
    \label{2.12}
    \hat{G}^\alpha = (C^{-1} )_{\alpha\beta} G^\beta.   
\end{equation}
The matrices $\hat{G}^\alpha$ are called \textit{cyclic}, if they  satisfy the \textit{cyclicity} condition
\begin{equation}
    \label{2.13}
    \hat{G}^\alpha_{\beta\gamma} = \hat{G}^\gamma_{\alpha\beta}\hspace*{6pt} \big(=  \hat{G}^\gamma_{\beta\alpha}\big), 
\end{equation}
the latter identity being satisfied by construction. If  $\hat{G}$ is cyclic, then the Gaussian measure with covariance \eqref{2.10} is time-stationary.

As for one component, the Gaussians shifted by a constant field are still time-stationary.  This is then equivalent to zero mean with an additional linear term, which is generically non-degenerate. To ensure a degenerate flux Jacobian, we therefore always assume zero mean for the Gaussian measure and also $A=0$, as before. Of interest is then the spacetime correlator
\begin{equation}
    \label{2.14}
    S_{\alpha\beta}(x,t) = \langle\phi_\alpha(x,t)\phi_\beta(0,0)\rangle.
\end{equation}
$S$ satisfies the sum rule
\begin{equation}
    \label{2.15}
    \int_\mathbb{R} dx S(x,t) = C.
\end{equation}
 
Repeating the scaling argument for a single component, one concludes that the dynamical exponent still equals $3/2$. Thus for each matrix element one would expect a similar scaling as in \equref{2.6}. But now no exact solution is of help and in our contribution we will barely scratch the surface. For sure, there are exceptional $\vec{G}$ matrices which lead to a different dynamical exponent. A trivial example would be $G^1_{11} = 1$ and all other matrix elements $0$. Then there will be a diffusive mode and a KPZ mode. Within the class of cyclic matrices we will formulate a conjecture which identifies those couplings having a dynamical exponent $3/2$.
\subsection{Time-invariance}
\label{2.3}
While one could study directly the continuum theory \cite{2015-funaki},
the argument becomes more transparent  by switching to the lattice, $j = 1,\ldots,N$, with periodic boundary conditions and denoting the discretized field by $\phi_{j,\alpha}$, $\alpha = 1,2$, see \cite{2009-sasamoto-spohn}. This discretization will also be used in our numerical simulations. The discretization of the linear part of \equref{2.8} is standard and reads 
\begin{equation}
    \label{2.16}
    \frac{d}{dt} \phi_{j,\alpha} 
    =  
    D_{\alpha\beta}(\phi_{j+1,\beta} -2\phi_{j,\beta}+\phi_{j-1,\beta}) + B_{\alpha\beta}(\xi_{j,\beta} - \xi_{j -1,\beta}).
\end{equation}
Condition  \eqref{2.11} is imposed and $\{\xi_{j,\beta}(t)\}$ is a family of independent standard white noises in time. The time-stationary measure for the linear Langevin equation is the Gaussian
\begin{equation}
    \label{2.17}
    \prod^N_{j=1} \prod^2_{\alpha'=1} \big((2\pi)^{-1} (\det C)^{-1/2}\exp\big[-\tfrac{1}{2}(C^{-1})_{\alpha\beta}\phi_{j,\alpha}\phi_{j,\beta}\big]  d \phi_{j,\alpha'}\big)= \rho_\mathrm{G} (\phi) \prod^N_{j=1} \prod^2_{\alpha'=1} d \phi_{j,\alpha'}.
\end{equation}
This measure should also be invariant under the deterministic nonlinear part of \equref{2.8} with $A = 0$. There is some freedom in the  discretization of nonlinearities. In our case, a natural constraint is to have the volume elements time-invariant, which is ensured by making the vector field divergence free, thereby leading to
\begin{equation}
    \label{2.18}
    \frac{d}{dt}\phi_{j,\alpha}= \tfrac{1}{3} G^\alpha_{\beta\gamma} \big(
    -\phi_{j-1,\beta}\phi_{j-1,\gamma}-  \phi_{j-1,\beta}\phi_{j,\gamma} + \phi_{j,\beta}\phi_{j+1,\gamma} + \phi_{j+1,\beta}\phi_{j+1,\gamma}\big).
\end{equation}
Then the time derivative of the action is 
\begin{eqnarray}
    \label{2.19}
    &&\hspace{0pt}\frac{d}{dt} \log\rho_\mathrm{G} =   -\tfrac{1}{3} \sum_{j=1}^N 
    \phi_{j,\alpha}(C^{-1})_{\alpha\beta}G^\beta_{\gamma\gamma'}\nonumber\\
    &&\hspace{60pt}\times \big(
    -\phi_{j-1,\gamma}\phi_{j-1,\gamma'}-  \phi_{j-1,\gamma}\phi_{j,\gamma'} + \phi_{j,\gamma}\phi_{j+1,\gamma'} + \phi_{j+1,\gamma}\phi_{j+1,\gamma'}\big)\nonumber\\
    &&\hspace{45pt}=   \tfrac{1}{3}\sum_{j=1}^N \big(\hat{G}^\alpha_{\gamma\gamma'} - \hat{G}^\gamma_{\alpha\gamma'}\big)( \phi_{j,\gamma}\phi_{j,\gamma'}\phi_{j+1,\alpha} -  \phi_{j,\alpha} \phi_{j+1,\gamma}\phi_{j+1,\gamma'}).
\end{eqnarray}
The sum vanishes only if the cyclicity condition \eqref{2.13} holds.

The cyclicity condition has been introduced in \cite{2014-spohn}.  From the perspective of stochastic analysis, cyclicity of $n$-component stochastic Burgers equations is studied in \cite{2015-funaki, 2023-hayashi,  2017-funaki-hoshino}.
\subsection{Cyclic matrices}
\label{sec2.4}
We assume that the fields have been linearly transformed such that  $C = 1$. In view of \equref{2.13}, the matrices $G^1$ and $G^2$ are cyclic, if they are of the form
\begin{equation}
    \label{2.20} 
    G^1 = 
    \begin{pmatrix}
        a & b\\
        b & c\\
    \end{pmatrix}, \qquad 
    G^2 = 
    \begin{pmatrix}
        b & c\\
        c & d\\
    \end{pmatrix}
\end{equation}
with arbitrary coefficients $a,b,c,d$. 

Since $C = 1$, it is natural to study how the dynamics behaves under orthogonal transformations of the fields. Reflection of either component yields the sign switches $(a,b,c,d) \rightarrow  (a,-b,c,-d)$ and $(a,b,c,d) \rightarrow  (-a,b,-c,d)$. Hence it suffices to consider the rotations given by  
\begin{equation}
    \label{2.21} 
    R = 
    \begin{pmatrix}
        \cos \vartheta & \sin \vartheta\\
        -\sin \vartheta& \cos \vartheta\\
    \end{pmatrix}
\end{equation}
with $\vartheta \in [0,2\pi]$, implying $R^\mathrm{T}R = 1$.  
Setting
\begin{equation}
    \label{2.22}
    \tilde{\phi}_\alpha = R_{\alpha\beta} \phi_\beta,
\end{equation}
the equations of motion 
\begin{equation}
    \label{2.23}
    \partial_t \phi_\alpha - \partial_x\big(\langle\vec{\phi},  G^\alpha \vec{\phi} \rangle+ D_{\alpha\beta}\partial_x\phi_\beta + B_{\alpha\beta}\xi_\beta\big) =0
\end{equation}
from \eqref{2.8} with $A=0$ transform to 
\begin{equation}
    \label{2.24}
    \partial_t \tilde{\phi}_\alpha - \partial_x\big(\langle\tilde{\vec{\phi}},  \tilde{G}^\alpha\tilde{\vec{\phi}} \rangle + \tilde{D}_{\alpha\beta}\partial_x\tilde{\phi}_\beta + \tilde{B}_{\alpha\beta}\xi_\beta\big) =0.
\end{equation}
The transformed matrices are obtained as
\begin{equation}
    \label{2.25}
    \tilde{D} = RDR^\mathrm{T}, \quad \tilde{B} = RBR^\mathrm{T},\quad \tilde{G}^\alpha = R_{\alpha\beta}RG^\beta R^\mathrm{T}.
\end{equation}
The matrices $\tilde{D},\tilde{B}$ are just another choice of $D,B$. 
In \ref{app1} we provide an explicit formula for $\tilde{G}^\alpha$, from which one verifies that the transformed matrices $\tilde{G}^1, \tilde{G}^2$ satisfy again the cyclicity condition with $C=1$. 
The transformed correlator is
\begin{equation}
    \label{2.26}
    \tilde{S}_{\alpha\beta}(x,t) = \langle\tilde{\phi}_\alpha(x,t)\tilde{\phi}_\beta(0,0)\rangle.
\end{equation}
Since the Gaussian measure with $C=1$ is invariant under rotations,  the correlator satisfies
\begin{equation}
    \label{2.27}
    \tilde{S}(x,t) = R\,S(x,t)R^\mathrm{T}.
\end{equation}
Thus the matrices $(\tilde{G}^1,\tilde{G}^2)$ are viewed as belonging to the same dynamical class as
$(G^1,G^2)$.
\subsection{Scaling hypothesis} 
\label{sec2.5}
As in the case of a single component, one argues that the transverse scale is $\xi_\bot \simeq t^{1/3}$ and the lateral scale $\xi_\| \simeq t^{2/3}$, see \cite{1990-krug-meakin, 1992-krug--healy}. Such property can be tested for specific observables. For the lateral scale,  specifically we consider  the spacetime correlator $S$ of \equref{2.14}. 
For it the scaling assumption states 
\begin{equation}
    \label{2.28}
    S(x,t) \simeq t^{-2/3}\mathsfit{g}( t^{-2/3}x)
\end{equation}
in approximation, for large $x,t$. 
The scaling function $\mathsfit{g}$ is a $2\times 2$ matrix-valued function which, in principle, depends  on all 
parameters of $\vec{G}$. According to the sum rule the normalization $\int dx \mathsfit{g}_{\alpha\beta}(x) = \delta_{\alpha\beta}$ holds.

To monitor the transverse scale the most common choice is to consider step initial conditions, which however requires to solve the Riemann problem for the coupled deterministic Burgers equations, i.e. upon omitting noise and diffusion. 
An attempt in this direction is reported in \cite{2016-mendl-spohn}. 
The more accessible option is either flat,  $\vec{\phi}(x) =  0$, or stationary initial conditions. The observable of interest is then the current across the origin
integrated over the time interval $[0,t]$, denoted by $\vec{\mathcal{J}}_0(t)$. It has a leading deterministic term proportional to $t$ and fluctuations on the transverse scale. Hence for large $t$,
\begin{equation}
    \label{2.29}
    \vec{\mathcal{J}}_0(t) \simeq \vec{v} t +  t^{1/3} \vec{\zeta}.
\end{equation}
$\vec{\zeta}$ is a random 2-vector whose distribution depends, in principle, on all 
parameters of $\vec{G}$.

 In case of a single component, clearly, at $\lambda = 0$ the scaling hypothesis fails and instead one
will observe Gaussian fluctuations. Within the cyclicity class, assuming $C=1$, the  scaling hypothesis fails for the choices  $(a,0,0,0)$ 
and $(0,0,0,d)$ and their rotations. Using \eqref{A.2} and \eqref{A.3}, in Dirac notation this leads to the exceptional coupling matrices 
\begin{eqnarray}
    \label{2.29a}
&&\hspace{100pt}G^1 = a \cos\vartheta|\cos\vartheta,-\sin\vartheta\rangle\langle \cos\vartheta,-\sin\vartheta|,\nonumber\\
&&\hspace{100pt}G^2 = -a \sin\vartheta|\cos\vartheta,-\sin\vartheta\rangle\langle \cos\vartheta,-\sin\vartheta|
\end{eqnarray}
and
\begin{eqnarray}
    \label{2.29b}
&&\hspace{100pt}G^1 = d  \sin\vartheta|\sin\vartheta,\cos\vartheta\rangle\langle \sin\vartheta,\cos\vartheta|,\nonumber\\
&&\hspace{100pt}G^2 = d \cos\vartheta |\sin\vartheta,\cos\vartheta\rangle\langle \sin\vartheta,\cos\vartheta|
\end{eqnarray}
for arbitrary $\vartheta \in [0,2\pi]$. As to be discussed, the hypothesis holds for $(0,b,0,0)$ with $b \neq 0$
and $(0,0,c,0)$  with $c\neq 0$. As a general experience from non-degenerate flux Jacobians, further increasing the number of non-zero couplings does not change the scaling behavior. Thus it seems to be safe to conjecture that the two cases listed above are the only exceptional parameter values. Away from cyclicity, no clear picture is yet available. The immediate obstacle is the structure of the steady state.
If the stationary measure has a finite correlation length, then anticipated is the same dynamical behavior as for cyclic couplings.
\subsection{Models to be studied} 
\label{sec2.6}
Since $C=1$, no basis for the fields is singled out. One choice of a canonical basis for $\vec{G}$ is then obtained through diagonalizing $G^2$, 
\begin{equation}
    \label{2.30} 
    G^1 = 
    \begin{pmatrix}
        a & b\\
        b & 0\\
    \end{pmatrix}, 
    \qquad 
    G^2 = 
    \begin{pmatrix}
        b & 0\\
        0 & d\\
    \end{pmatrix}.
\end{equation}
In fact, for our numerical studies we will choose $a = 0$, so to be able to compare with the results of \cite{2023-nardis--vasseur}. Thus
\begin{equation}
    \label{2.31} 
    G^1 = 
    b \begin{pmatrix}
        0 & 1\\
        1 & 0\\
    \end{pmatrix}, 
    \qquad 
    G^2 = 
    b\begin{pmatrix}
        1 & 0\\
        0 & \lambda\\
    \end{pmatrix},
\end{equation}
where $\lambda$ is regarded as a free parameter and $b$ can be absorbed into a modified time scale. 

The $G^2_{11}$ matrix element fixes the sign and hence $\lambda \in \mathbb{R}$. For $\lambda = 1$, the commutator 
$[G^1,G^2]= 0$ and through a  rotation of $\pi/4$, the matrices are transformed to 
\begin{equation}
    \label{2.32} 
    \tilde{G}^1 = 
    \sqrt{2}b \begin{pmatrix}
        1 & 0\\
        0 & 0\\
    \end{pmatrix}, 
    \qquad 
    \tilde{G}^2 = 
    \sqrt{2}b  \begin{pmatrix}
        0& 0\\
        0 & 1\\
    \end{pmatrix}.
\end{equation}
Hence the system decouples into two one-component stochastic Burgers equations.


\section{Numerical simulations of coupled stochastic Burgers and KPZ equations}
\label{sec3}
\setcounter{equation}{0}

\begin{figure}[h]
    \begin{subfigure}{0.5\linewidth}
	\includegraphics[width=1\linewidth]{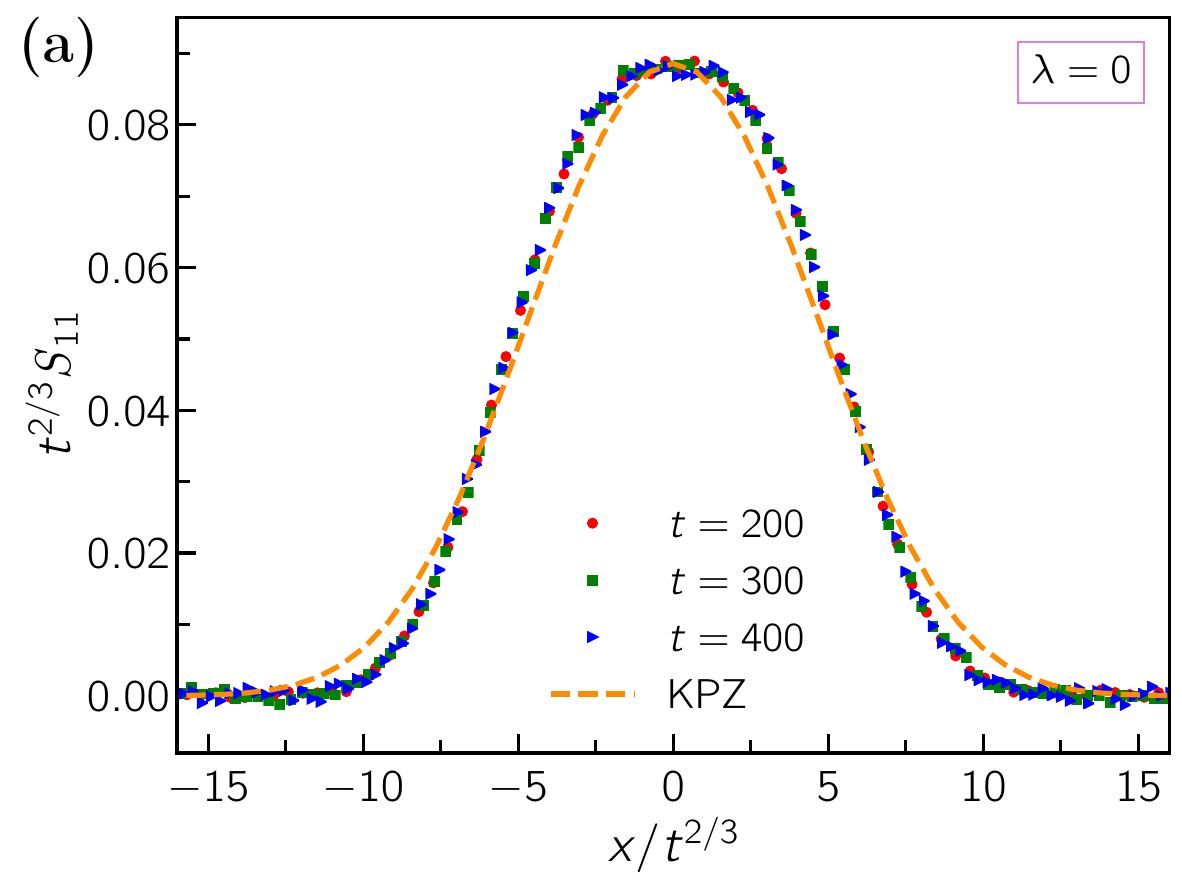}
    \end{subfigure}%
    \begin{subfigure}{0.5\linewidth}
	\includegraphics[width=1\linewidth]{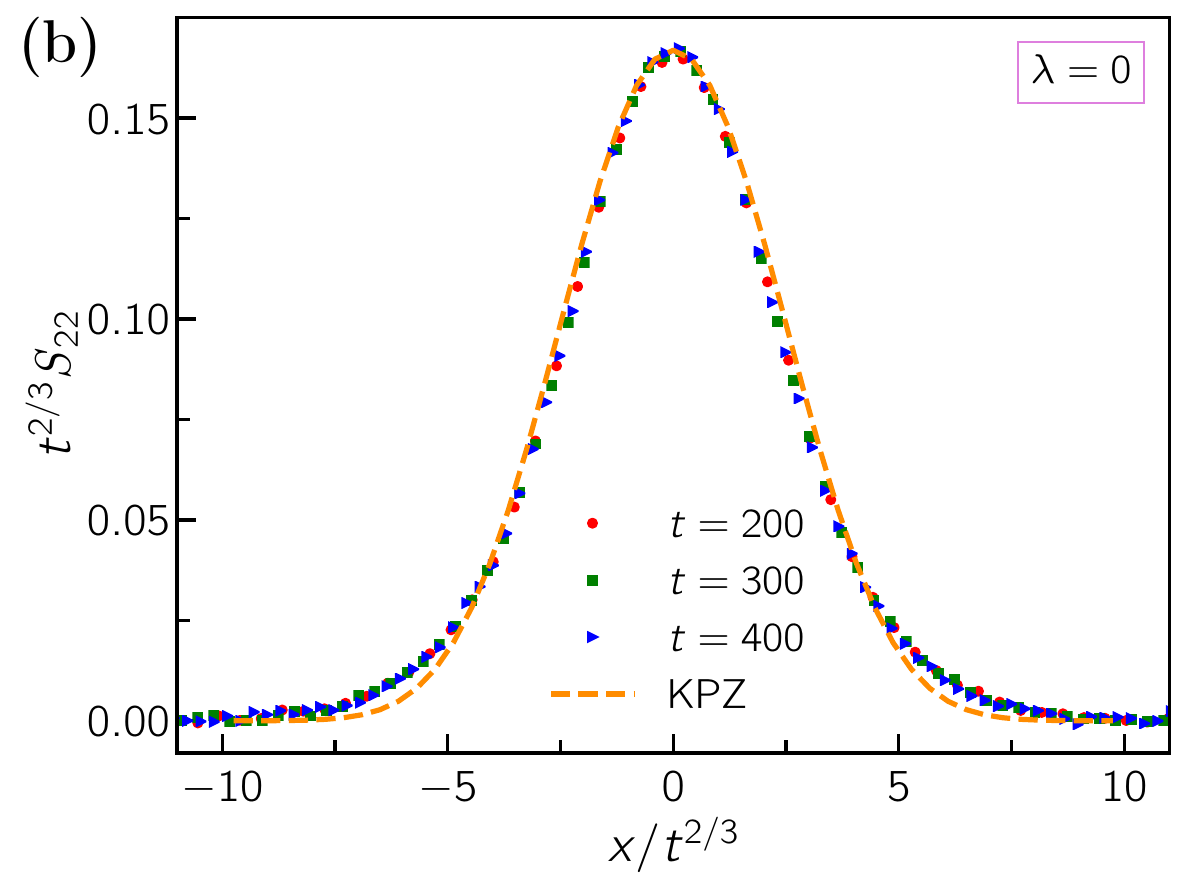}
    \end{subfigure}
    \caption{Correlators $S_{11}$ and $S_{22}$ obtained from numerical simulations of the coupled stochastic Burgers equations in \equref{3.1} with $b=3$ and $\lambda=0$ using the discretizations of Eqs.~\eqref{2.16} and \eqref{2.18}. The system size is $L=4096$ and the number of independent simulations is $10^5$. The covariance of the initial correlator is $\langle \phi_{\alpha,j}\phi_{\beta,0} \rangle = \delta_{\alpha\beta}\delta_{j0}$. The sum rule, $\sum_j S_{11}(j,t) = \sum_j S_{22 }(j,t) = 1$, is satisfied within a numerical accuracy of $1\%$. 
    }
    \label{fig:1}
\end{figure}

For the numerical studies, we focus on the model defined by the  matrices $G^1,G^2$
of \equref{2.31}. Hence the Langevin equations are
\begin{equation}
    \begin{aligned}
        \partial_t \phi_1 &= \partial_x \big(  \partial_x \phi_1 + 2 b \phi_1 \phi_2 + \sqrt{2}\xi_1 \big), \\
        \partial_t \phi_2 &= \partial_x \big( \partial_x \phi_2 + b (\phi_1^ 2 + \lambda \phi_2^2)+ \sqrt{2}\xi_2 \big).
    \end{aligned}
    \label{3.1}
\end{equation}
In our simulations, diffusion is fixed to $1$ and noise strength to $\sqrt{2}$, so that  the static covariance equals $C=1$. The initial data are random and given by the Gaussian measure \eqref{2.17} with $C=1$. 
For $\lambda =1$, after a rotation by $\pi/4$, the system decouples into two independent stochastic Burgers equations. For our numerical simulations we focus on the case $\lambda=0$. For a non-degenerate
Jacobian, only by setting some coefficients to $0$ a distinct universality can be accomplished. Thus $\lambda = 0$ has a good chance for exhibiting a non-KPZ  behavior. We perform direct numerical simulations (DNSs) of \equref{3.1} using the discretization scheme provided in Eqs.~\eqref{2.16}, \eqref{2.18}. Of interest is the spacetime correlator $S_{\alpha\beta}$ as defined in \equref{2.14}. The Gaussian measure is invariant under  the change $\phi_1 \leadsto  - \phi_1$. Invariance of ~\equref{3.1} under this transformation implies that $\phi_1(x,t)$ has an even distribution.
From this property one concludes that $S_{12}(x,t) = 0  =
S_{21}(x,t)$, which is very well satisfied numerically.

The real task is to compute $S_{11}$ and $S_{22}$ for $\lambda =0$. Our results are displayed in \figref{fig:1} and,
considering the full time sequence (not plotted), it appears
that the numerical results are already in the scaling regime.
First of all, one notes that the anticipated $t^\frac{2}{3}$ scaling exponent is very well satisfied. At $\lambda = 1$, the system is decoupled and $S_{11}$, $S_{22}$ would scale with $f_\mathrm{KPZ}$ having the same value for coefficient $\Gamma_{\|}$.
At $\lambda=0$ one observes that $S_{11}$ is bulky and $S_{22}$ slim. More importantly they differ from 
$f_\mathrm{KPZ}$. Even allowing for linear scale change with a single parameter $\Lambda$, $S_{11}$ differs significantly from $S_{22}$. 
Thus as, a so far preliminary, conclusion: the scaling exponent conforms with the one-component model, but the scaling function  depends on the coupling $\lambda$ not merely through a single non-universal time scale coefficient $\Gamma_{\|}$.

\begin{figure}[h]
    \includegraphics[width=1\linewidth]{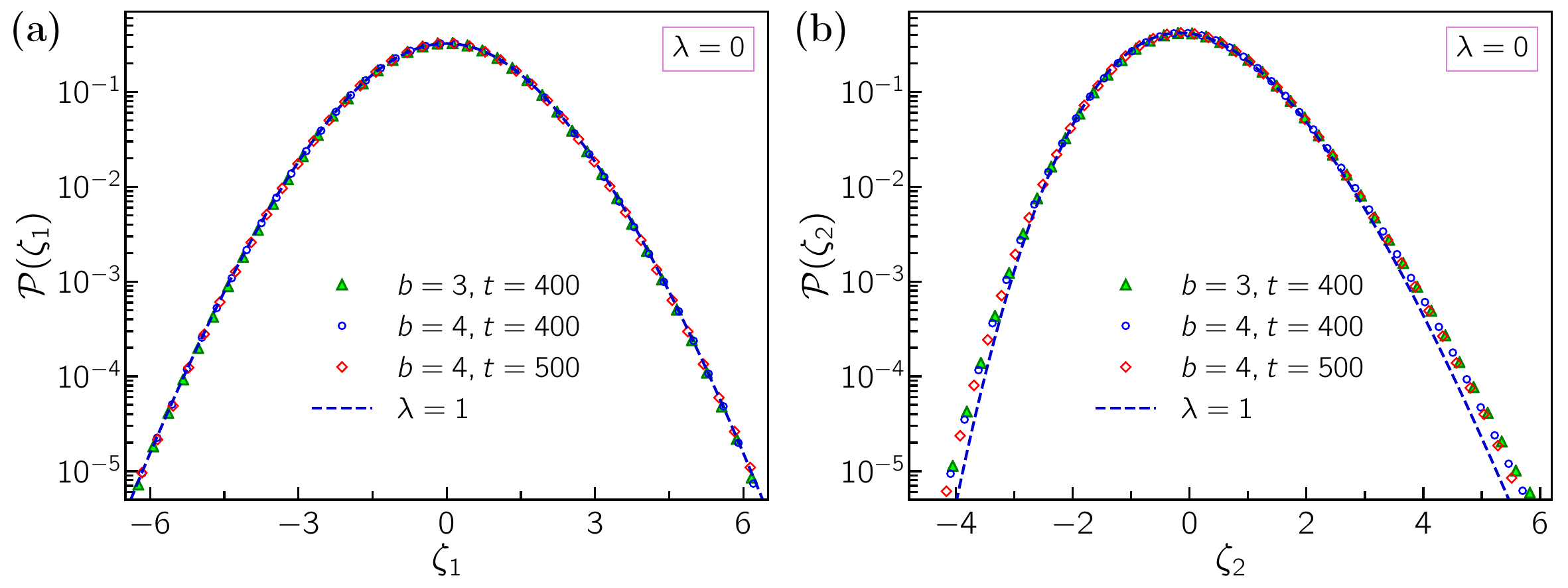}
    \caption{Probability density functions (PDFs) of (a) $\zeta_1$ and (b) $\zeta_2$ for the coupled KPZ equations \eqref{3.3} at $\lambda=0$ for different values of the parameter $b$ and a sequence of times $t$. The system size is $L=8192$ with
    periodic boundary conditions. The total number of simulation runs is $5\times 10^4$. For both $\zeta_1$ and $\zeta_2$ we set $\Gamma_\perp=b$.  The corresponding PDFs collapse on a single curve. The variances for $\zeta_1$ and $\zeta_2$ are $1.545$ and $0.943$, respectively. Comparison is with the corresponding quantities at $\lambda=1$. } 
    \label{fig:2}
\end{figure}

The second task concerns the time integrated currents. By Eq.~\eqref{3.1} the instantaneous currents are defined as 
\begin{equation}
    \label{3.1a}
\mathcal{J}_1
= -\big(\partial_x \phi_1 + 2 b \phi_1 \phi_2 + \sqrt{2}\xi_1\big),\qquad
\mathcal{J}_2 = -\big(\partial_x \phi_2 + b (\phi_1^ 2 + \lambda \phi_2^2)+ \sqrt{2}\xi_2\big).
\end{equation}
Numerically, it is convenient to switch to the height field through
\begin{equation}
    \label{3.2}
    \pd{x} h_\alpha = \phi_\alpha.
\end{equation}
The Langevin equations for the KPZ height fields $h_\alpha$ are then
\begin{equation}
    \label{3.3}
    \begin{aligned}
	\pdt h_1 &= \pd{xx} h_1 + 2 b \, \pd{x} h_1 \, \pd{x} h_2 + \sqrt{2}\xi_1  , \\
	\pdt h_2 &= \pd{xx} h_2 + b \big( (  \pd{x} h_1 )^2 + \lambda ( \pd{x} h_2 )^2 \big)+ \sqrt{2}\xi_2 .
    \end{aligned}
\end{equation}
Because of the conservation law, height fields and currents are related as 
\begin{equation}
    \label{3.4}
    h_\alpha(x,t) - h_\alpha(x,0) = -\int_0^t ds \mathcal{J}_\alpha(x,s),
\end{equation}
where $\mathcal{J}_\alpha(x,s)$ is the instantaneous current \eqref{3.1a} evaluated at $(x,s)$. For the simulations, space is 
discretized with index $0,...,L-1$ and the height satisfies helical  boundary conditions. In our set up,
the initial height field has
independent unit Gaussian zero mean increments and helical constraint becomes $h_L = h_0$. 
At large  times, to leading order the height fields increase linearly in $t$,
\begin{equation}
    \label{3.5}
    h_{\alpha}(t) \simeq v_{\alpha}t.
\end{equation}
For the discretization according to Eqs.~\eqref{2.16}, \eqref{2.18} the velocity is given by
\begin{equation}
    \label{3.5}
    v_{\alpha} =\tfrac{2}{3} G^\alpha_{\gamma\gamma},
\end{equation}
as well confirmed numerically. To study the fluctuations, the linear term is subtracted leading to 
\begin{equation}
   \zeta_\alpha = \Gamma_\perp^{-\frac{1}{3}} t^{-1/3} \big(h_{\alpha}(x,t) -h_{\alpha}(0) -v_{\alpha} t \big),
     \label{3.6}
\end{equation}
where for notational simplicity we introduced already the scale factor $\Gamma_\perp^{-\frac{1}{3}}$.

In \figref{fig:2}, we confirm that $\zeta_\alpha$ has a well-defined limiting distribution, independent of $x$ by stationarity. By definition $\vec{\zeta}$ is a random vector, which has a two-dimensional distribution.
Numerically determined are only the two marginals.  In \figref{fig:2} we display the PDFs of  $\zeta_1$ and $\zeta_2$ at $\lambda = 0$ and compare with the corresponding PDFs at $\lambda=1$.
Note that at $\lambda= 1$, after a rotation  by $\pi/4$, the resulting vector $(\zeta_+,\zeta_- ) = (1/\sqrt{2})(\zeta_1 +\zeta_2,- \zeta_1 +\zeta_2)$ has independent components, both  Baik-Rains distributed \cite{2000-baik-rains}. Hence in \figref{fig:2} we compare with sum and differences of two independent Baik-Rains PDFs. As we decrease $\lambda$, independence is lost and the probability density functions of the two marginals appear to  differ significantly from the one of Baik-Rains. Note that $\zeta_-$ is symmetric but distinct from a Gaussian. 

\begin{table}[h]
    \begin{center}
    \renewcommand*{\arraystretch}{2}
    \begin{tabular}{|c|c|c||c|c|}
	\hline
	Distribution & $\mathbbm{M}$ & $\mathbbm{V}$ &  $\mathbbm{S}$  & $\mathbbm{K}$      
        \\
	\hline
	Baik-Rains \cite{2000-baik-rains, 2000-prahofer-spohn} & $0$ & $1.150$ &          $0.359$ & $0.289$ 
        \\
	\hline
	$ \zeta_1 $ & $0$ & $1.545$ & $0$ & $0.152$ 
        \\
	\hline
	$\zeta_2$ & $-0.025$ & $0.943$ & $0.250$ & $0.236$ 
        \\
	\hline
        $\brac{ \zeta_1 + \zeta_2}/\sqrt{2}$ & $-0.018$ & $1.243$ & $0.307$ & $0.253$
        \\
        \hline
        $\brac{ -\zeta_1 + \zeta_2 }/\sqrt{2}$ & $-0.017$ & $1.245$ & $0.307$ & $0.256$ 
        \\
        \hline
    \end{tabular}
    \caption{Comparing skewness $\mathbbm{S}$ and excess kurtosis $\mathbbm{K}$ of the Baik-Rains distribution with the ones for $\zeta_1$ and $\zeta_2$. These values are obtained by averaging over the time interval $[425, 500]$ for $b=4$ and $L=8192$. By stationarity, the means of $\zeta_\alpha$ equal $0$. The last two rows display the values for symmetric and antisymmetric combinations of $\zeta_1$ and $\zeta_2$, see also \ref{sec:pm}.}
    \label{tab:1}
    \end{center}
\end{table}

For a more quantitative comparison, the dimensionless form of skewness, $\mathbbm{S}$, and kurtosis, $\mathbbm{K}$ is listed in Table \ref{tab:1}, where for some random variable, $X$, with mean $\mathbbm{M}$ and variance $\mathbbm{V}$, skewness and kurtosis are  defined through
\begin{equation}
    \mathbbm{M} = \langle X \rangle , \ \ 
    \mathbbm{V} = \langle ( X - \mathbbm{M} )^2 \rangle, \ \ 
    \mathbbm{S} = \mathbbm{V}^{-3/2}\langle  (X - \mathbbm{M} )^3 \rangle , \ \
    \mathbbm{K} = \mathbbm{V}^{-2}\langle  (X - \mathbbm{M} )^4 \rangle - 3. 
    \label{3.8}
\end{equation}
 We also added the values for $(\pm\zeta_1 + \zeta_2)/\sqrt{2}$ (see \ref{sec:pm}). 
$\zeta_1$ and $\zeta_2$ are not independent and the sum is performed for each realization. As to be noted, in terms of skewness and kurtosis, $\zeta_1$ differs significantly from Baik-Rains, while $\zeta_2$ is still different,
but somewhat less so.


\section{Two-lane lattice gas}
\label{sec4}
\setcounter{equation}{0}

For one component, KPZ universality is confirmed not only for the continuum equation, but also for discrete models as  TASEP and ASEP. Such results strongly support the KPZ scaling theory which asserts the lateral time scale $t^\frac{2}{3}$, the transversal time scale $t^\frac{1}{3}$, universal scaling functions, and corresponding non-universal coefficients. Up to constants independent of $\lambda$ , the latter are given by $\Gamma_\| = 2|\lambda| \sqrt{2\chi}$ and $\Gamma_\bot = |\lambda|\chi^2$, where $\lambda = j''(\rho)/2$ with $j(\rho)$ the average current.  Examples of  two-component systems with a non-degenerate flux Jacobian are the Arndt-Heinzel-Rittenberg (AHR) \cite{1998-arndt--rittenberg, 1999-arndt--rittenberg, 2013-ferrari--spohn} and the Bernardin-Stoltz (BS) \cite{2012-bernardin-stoltz, 2015-spohn-stoltz}
models. The predictions based on nonlinear fluctuating hydrodynamics \cite{2012-beijeren, 2014-spohn} have been well confirmed in \cite{2013-ferrari--spohn} and \cite{2015-spohn-stoltz}, but do not cover the degenerate case.

The first task is to find a suitable lattice gas model. Our search has been guided by possessing a product invariant stationary measure and a minimal number of states at each lattice site. The AHR model has three local states, (-,0,+). For its stationary measures the matrix  product construction is required \cite{2013-ferrari--spohn}. As it turns out, the flux Jacobian is non-degenerate over the entire parameter range. Popkov and Sch\"{u}tz studied two-lane models with local states $(00,01,10,11)$.
The version studied in \cite{2014-popkov--schutz} satisfies our conditions, but the flux Jacobian is again non-degenerate. An earlier variant can be found in \cite{2003-popkov-schutz}. This two-lane model allows for degeneracy, as investigated in \cite{2012-popkov-schutz} under the label ``umbilic point''. The flux Jacobian is still explicit. Hence this stochastic two-lane lattice gas is optimally qualified for our purposes.

The basic elements of the KPZ scaling theory are easily generalized. The lattice gas has two conserved densities. The steady states are labelled by the average densities $\vec{\rho}$. For our model  the average currents, $\vec{j}(\vec{\rho})$, are given and hence also the flux Jacobian. $(G^1,G^2)$  are the matrices of second derivatives for $j_1(\vec{\rho}),j_2(\vec{\rho})$. The unique point of degeneracy turns out to be density $1/2$ in each lane. To compare with \equref{3.1}, the crucial insight is to first match the static susceptibility. For the continuum equation we adopted the convention $C=1$. Thus for the lattice gas we have to diagonalize $C$ and then rescale so to achieve a unit matrix. This linear transformation has to be applied to $(G^1,G^2)$, as already discussed in Section 2. The transformed matrices then appear in the approximation through a stochastic continuum equation.

\begin{figure}[h]
    \begin{center}
	\includegraphics[width=0.9\linewidth]{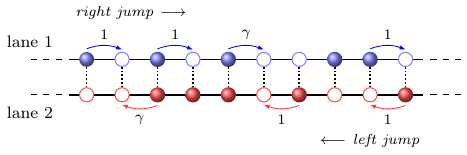}
    \end{center}
    \caption{Illustrating the two-lane lattice gas model \cite{2012-popkov-schutz}. The lattice sites are indexed by $(\alpha, n)$ where $\alpha=1, 2$, $n \in \mathbb{Z}$.} 
    \label{fig:3}
\end{figure}

In the two-lane model of Popkov and Sch\"{u}tz, the sites are labeled by $(\alpha, n)$, $\alpha = 1,2$,
$n \in\mathbb{Z}$. The occupation variables are $\eta_{\alpha,n}$ with values $0,1$,  where $\alpha= 1$
is also referred to as top lane and $\alpha = 2$ as bottom lane, see \figref{fig:3}. Under the exclusion rule, particles in lane 1 jump only to the right and in lane 2 only to the left. The jump rates, equivalently exchange rates, are given by
\begin{equation}
    \label{4.1}
    \begin{aligned}
        W_{n,n+1}^{1} &= \eta_{1,n} (1-\eta_{1,n+1}) + (\gamma -1) \eta_{1,n} (1-\eta_{1,n+1}) \eta_{2,n} (1-\eta_{2,n+1}), \\
        W_{n, n+1}^{2} &= (1-\eta_{2,n})\eta_{2,n+1}  + (\gamma -1) (1-\eta_{2,n})\eta_{2,n+1}  (1-\eta_{1,n})\eta_{1,n+1}  
    \end{aligned}
\end{equation}
with $\gamma >0$. Particles stay in their own lane. If $\gamma = 1$, then lane 1 is TASEP with jumps of rate $1$ to the right and lane 2 is TASEP with jumps of rate 1 to the left. The interaction between lanes results from varying $\gamma$. If in lane $1$ at sites $n,n+1$ the configuration  is $10$, then at sites $n,n+1$ in lane 2  the possible configurations are $00, 10,01,11$. In case $10$ shows, the jump rate in lane 1 is modified to $\gamma$, while in the three other cases the jump rate stays  to be equal to 1. If in lane $2$ at sites $n,n+1$ the configuration  is $01$, the jump rates are given by the corresponding mirror image, see \figref{fig:3}.

As established in \cite{2003-popkov-schutz}, the invariant measures are product with respect to $n$. To study their properties it thus suffices to consider a single site. For lighter notation we set $\eta_{1,0} = \vartheta$ and $\eta_{2,0}=\theta$. Then the possible states are $00,10,01,11$ and the single site measure of the steady state is
\begin{equation}
    \label{4.2}
    p_{ \vartheta\theta }= \frac{1}{Z} \exp\big(-2\nu \vartheta \theta +\mu \vartheta+\tilde{\mu} \theta \big).
\end{equation}
Here $\gamma = \mathrm{e}^{2\nu}$, $\mu,\tilde{\mu} $ are chemical potentials, and $Z$ is normalizing partition function. (Note that our $2\nu$  equals $\nu$ of \cite{2012-popkov-schutz}). Setting $p_{11} = \Omega$, one obtains
\begin{equation}
    \label{4.3}
    \Omega =\frac{1}{2q}\Big( -1- q(1-u-v) + \sqrt{(1+q(1-u-v))^2 +4quv }\Big)
\end{equation}
with
\begin{equation}
    \label{4.4}
    q = \gamma - 1,\quad u = \langle\vartheta\rangle,\quad v = \langle \theta\rangle.
\end{equation}
The particle density on the top-lane is $u$ and the one on the bottom-lane $v$. Then
\begin{equation}
    \label{4.5}
    p_{00} = 1 -u-v +\Omega, \quad p_{01} = v - \Omega,\quad p_{10} = u- \Omega, \quad p_{11}  = \Omega. 
\end{equation}
The static covariance matrix is given by
\begin{equation}
    \label{4.6}
    C_{\alpha\beta}(u,v)= \langle \eta_{\alpha,0}\eta_{\beta,0} \rangle - 
    \langle \eta_{\alpha,0} \rangle \langle \eta_{\beta,0} \rangle,\qquad
    C(u,v) = 
    \begin{pmatrix}
        u(1-u) & \Omega - uv\\
        \Omega - uv\ & v(1-v)\\
    \end{pmatrix}.
\end{equation}
The average currents have been computed in \cite{2003-popkov-schutz, 2012-popkov-schutz}
with the result
\begin{equation}
    \label{4.7}
    j_1(u,v) = u(1-u) +q\Omega(1-u-v +\Omega), \qquad j_2(u,v) = - v(1-v) -q\Omega(1-u-v +\Omega).
\end{equation}

From \equref{4.7} one deduces the flux Jacobian 
\begin{equation}
    \label{4.8}
    A= \begin{pmatrix}  
        1-2u & 0\\
        0 & -1+2v\\
    \end{pmatrix}
    +q\Omega \begin{pmatrix}  
        -1 & -1\\
        1 & 1\\
    \end{pmatrix} 
    + q\big((1-u-v) +2\Omega\big)\begin{pmatrix}  
        \partial_u \Omega&  \partial_v \Omega\\
         -\partial_u \Omega &  -\partial_v \Omega\\
    \end{pmatrix} . 
\end{equation}
For $q=0$, the line of degeneracy equals $\{1 - u-v = 0\}$. As proved in the Appendix, for 
$q > -1$, $q \neq 0$, the only point of degeneracy in the interior of the square 
$[0,1]^2$  is $u = \tfrac{1}{2} = v$, in which case both eigenvalues are $0$. 
Then the covariance simplifies to
\begin{equation}
    \label{4.9}
    C(\tfrac{1}{2},\tfrac{1}{2}) = C = 
    \tfrac{1}{4}\begin{pmatrix}  
        1 & -1\\
        -1 & 1\\
    \end{pmatrix} + \Omega
    \begin{pmatrix}
        0 & 1\\
        1 & 0\\
    \end{pmatrix},
\end{equation}
which has eigenvalues  
\begin{equation}
    \label{4.10}
    \lambda_1^2 = \Omega(\tfrac{1}{2},\tfrac{1}{2}) = \frac{1}{2(1 + \mathrm{e}^{\nu})}, \qquad \lambda_2^2 =  \tfrac{1}{2}- \Omega(\tfrac{1}{2},\tfrac{1}{2})
    =\frac{1}{2(1 + \mathrm{e}^{-\nu})}.
\end{equation}
The average currents turn out to be given by
\begin{equation}
    \label{4.11}
    j_1(\tfrac{1}{2},\tfrac{1}{2}) = \frac{1}{2(1 + \mathrm{e}^{-\nu})},
    \qquad j_2(\tfrac{1}{2},\tfrac{1}{2}) = 
    - j_1(\tfrac{1}{2},\tfrac{1}{2}).
\end{equation}

 To relate the lattice gas with the continuum Burgers equation in \eqref{2.8}, we follow the strategy laid out in nonlinear fluctuating hydrodynamics. The currents \eqref {4.7}  determine the Euler equations of the two-lane model, which can be written as 
\begin{equation}
\label{4.11a}
\partial_t u +\partial_x j_1(u,v) = 0,\qquad
\partial_t v +\partial_x j_2(u,v) = 0.
\end{equation}
    The background state has constant densities $\tfrac{1}{2}$. Thus we expand the currents $j_1,j_2$ as $u = \tfrac{1}{2} + u_1$, $v = \tfrac{1}{2} + u_2$. The linear term is the flux Jacobian evaluated at $u = \tfrac{1}{2} = v$. This term vanishes, thereby confirming that in the  steady state there is no ballistic component. To second order in $u_1,u_2$ one obtains 
\begin{eqnarray}\label{4.11b}
\hspace{90pt} j_1(\tfrac{1}{2} + u_1,\tfrac{1}{2} + u_2) &= & 
j_1(\tfrac{1}{2},\tfrac{1}{2}) + \tfrac{1}{2}\langle\vec{u},H^1 \vec{u}\rangle, \nonumber\\
\hspace{90pt} j_2(\tfrac{1}{2} + u_1,\tfrac{1}{2} + u_2)& = & 
-j_1(\tfrac{1}{2},\tfrac{1}{2}) + \tfrac{1}{2}\langle\vec{u},H^2 \vec{u}\rangle.
  \end{eqnarray}
Here the second order matrices are 
 \begin{equation}
    \label{4.12}
    H^1 = 
    \begin{pmatrix}  
        -1 & 0\\
        0 & 0\\
    \end{pmatrix} 
    -H^{\diamond}, \qquad 
    H^2 = 
    \begin{pmatrix}  
        0 & 0\\
        0 & 1\\
    \end{pmatrix} 
    + H^{\diamond}
\end{equation}
with
\begin{equation}
    \label{4.13}
    H^{\diamond} = 
    \tfrac{1}{2}\begin{pmatrix}  
        \sinh\nu & \cosh \nu -1\\
        \cosh \nu -1 & \sinh\nu\\
    \end{pmatrix}. 
\end{equation}

Recalling the discussion in Section 2,  a comparison with the continuum equations requires to transform  the static covariance $C$ to the identity matrix. This is 
done in the two steps $\tilde{u}_\alpha = R_{\alpha\beta} u_\beta$ and $\varphi_\alpha = (Q^{-1})_{\alpha\beta}\tilde{u}_\beta$, 
where $R$ represents rotation by $\pi/4$, 
\begin{equation}
    \label{4.15}
    R = 
    \frac{1}{\sqrt{2}}\begin{pmatrix}  
        1 & 1\\
        - 1& 1\\
    \end{pmatrix},
\end{equation}
and $Q$ is the diagonal matrix
\begin{equation}
    \label{4.16}
    Q = 
    \begin{pmatrix}  
        \lambda_1 & 0\\
        0&\lambda_2\\
    \end{pmatrix}.
\end{equation}
Hence $C$ is transformed as
\begin{equation}
    \label{4.14}
    Q^{-1}RCR^\mathrm{T} Q^{-1} = 1.
\end{equation}

Applying the rotation to $H^\alpha$ yields
\begin{equation}
    \label{4.17}
    2RH^1R^\mathrm{T} =
    \begin{pmatrix}  
        -1 & 1\\
        1& -1\\
    \end{pmatrix}
    -\begin{pmatrix}  
        \mathrm{e}^\nu -1& 0\\
        0& 1 - \mathrm{e}^{-\nu}\\
    \end{pmatrix}
\end{equation}
and
\begin{equation}
    \label{4.18}
    2RH^2R^\mathrm{T} =
     \begin{pmatrix}  
        1 & 1\\
        1& 1\\
    \end{pmatrix}
    +\begin{pmatrix}  
        \mathrm{e}^\nu -1& 0\\
        0&1 - \mathrm{e}^{-\nu}\\
    \end{pmatrix}.
\end{equation}
Using $\tilde{H}^\alpha =R_{\alpha\beta}RH^\beta R^\mathrm{T}$, valid for orthogonal transformations, one arrives at
\begin{equation}
    \label{4.19}
    \tilde{H}^1=
    \frac{1}{ \sqrt{2} } 
    \begin{pmatrix}  
        0 & 1\\
        1& 0\\
    \end{pmatrix} ,\qquad
    \tilde{H}^2=
    \frac{1}{ \sqrt{2} } 
    \begin{pmatrix}  
        \mathrm{e}^\nu& 0\\
        0& 2 - \mathrm{e}^{-\nu}\\
    \end{pmatrix} .
\end{equation}
In the second step  $\tilde{H}^\alpha$  is transformed under $Q^{-1}$. We denote the transformed matrices by $G^\alpha$,
since they appear in the continuum approximation \eqref{2.8} with $A=0$ and $C=1$,
i.e. $D + D^\mathrm{T} = BB^\mathrm{T}$. In formulas,
because $Q = Q^\mathrm{T} > 0$, one has 
$G^\alpha= (Q^{-1})_{\alpha\beta}Q\tilde{H}^\beta Q$ and hence
\begin{equation}
    \label{4.20}
    G^1=
    b\begin{pmatrix}  
        0 & 1\\
        1& 0\\
    \end{pmatrix} ,\qquad
    G^2=
     b\begin{pmatrix}  
        1& 0\\
        0& 2 - \frac{1}{\sqrt{\gamma}}
    \end{pmatrix}, \quad b = \frac{1}{2 \sqrt{2}}\lambda_2.
\end{equation}
The extra factor of $1/2$ in the definition of $b$ results from the Taylor expansion in 
\eqref{4.11b}. Evidently $\vec{G}$ is cyclic. In \equref{4.20} the parameter $b$ is merely a time scale. Multiplying all jump rates with a constant factor would modify the parameter $b$. More important is the relation  
\begin{equation}
\label{5.20a}
\lambda = 2 - \frac{1}{\sqrt{\gamma}},
\end{equation}
which determines the parameter of the stochastic Burgers equation at which its
asymptotic behavior coincides with the one of the lattice gas.

Since $\gamma > 0$, the range of $\lambda$ is $[-\infty,2]$, $\lambda = 0$
corresponding to $\gamma = 0.25$. The lattice gas cannot reach values of $\lambda$ beyond $2$. At $\gamma = \infty$, the lattice gas  is still well defined, only some specific exchanges occur instantaneously. Surprisingly, also negative values of $\lambda$ show 
 up. Note that the sign of $\lambda$ cannot be trivially removed due to $b^{-1}G^2_{11} = 1$. For small $\gamma$, there are many almost blocked configurations and the current vanishes as
 $\sqrt{\gamma}$. This
 indicates that the convergence to the scaling regime will be very slow.

These results could not have been anticipated. The strategy employed is general, but 
the fact that the two-lane model turns out to have cyclic $\vec{G}$ matrices is specific.
Nonlinear fluctuation theory predicts that on large scales the two-lane model
and the stochastic Burgers equation \eqref{3.1} with $\lambda = 2 - 1/\sqrt{\gamma}$  and $b =  \lambda_2/2 \sqrt{2} $ have the same scaling behavior,
provided the fields are properly transformed.
A numerical check will be discussed in Section \ref{sec6}.
In the case of a non-degenerate flux Jacobian, according to nonlinear fluctuating hydrodynamics the correctly adjusted fields are the eigenmodes of $A$. In the degenerate case these turn trivial and instead the covariance matrix $C$ plays a central role.


\begin{figure}[h]
    \begin{subfigure}{0.5\linewidth}
	\includegraphics[width=1\linewidth]{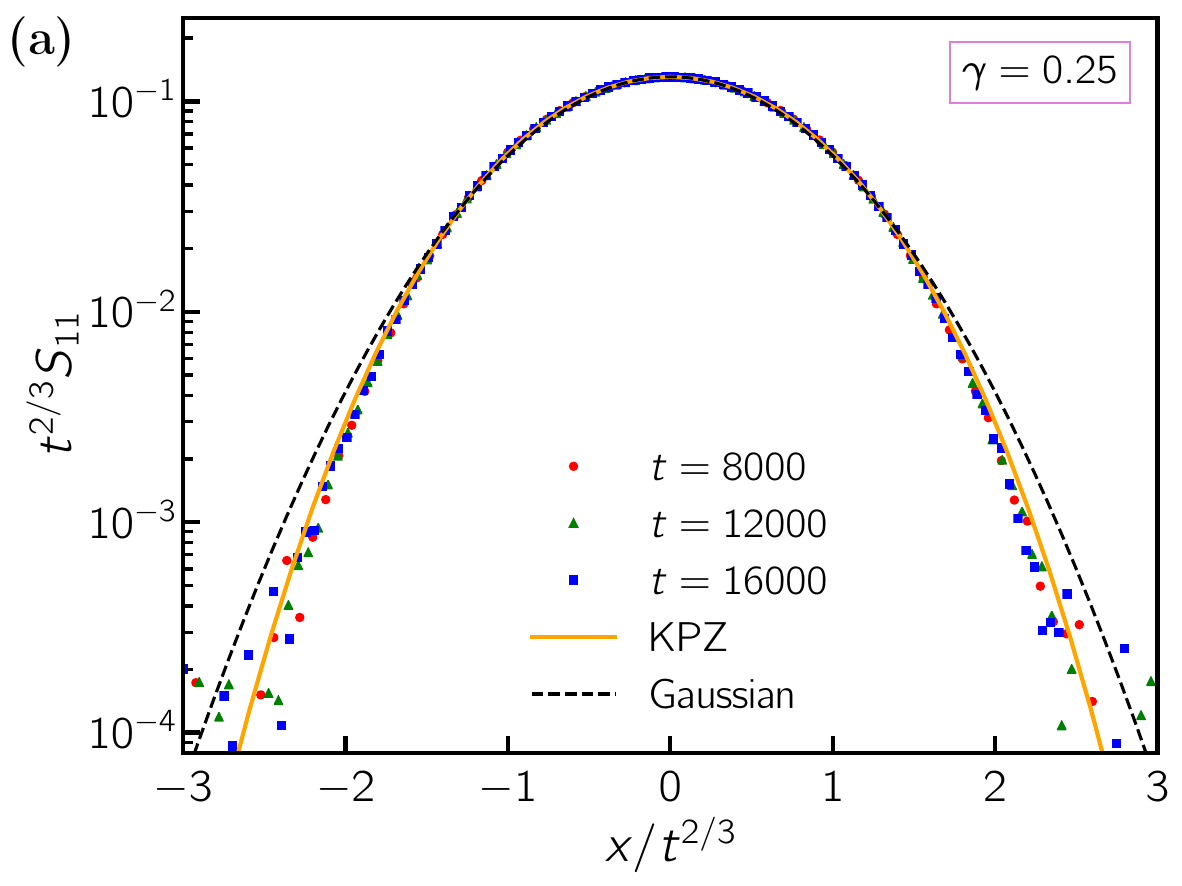}
    \end{subfigure}%
    \begin{subfigure}{0.5\linewidth}
	\includegraphics[width=1\linewidth]{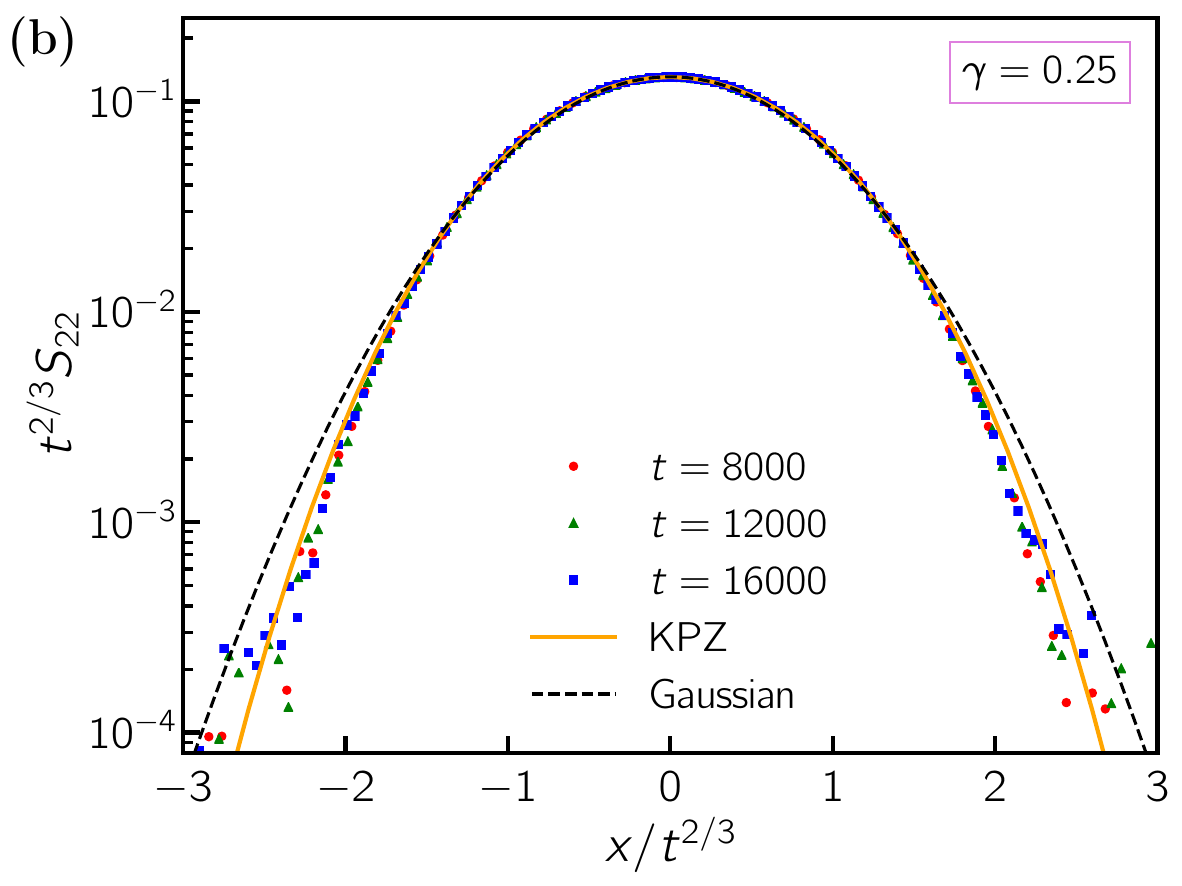}
    \end{subfigure}
    \begin{subfigure}{0.5\linewidth}
	\includegraphics[width=1\linewidth]{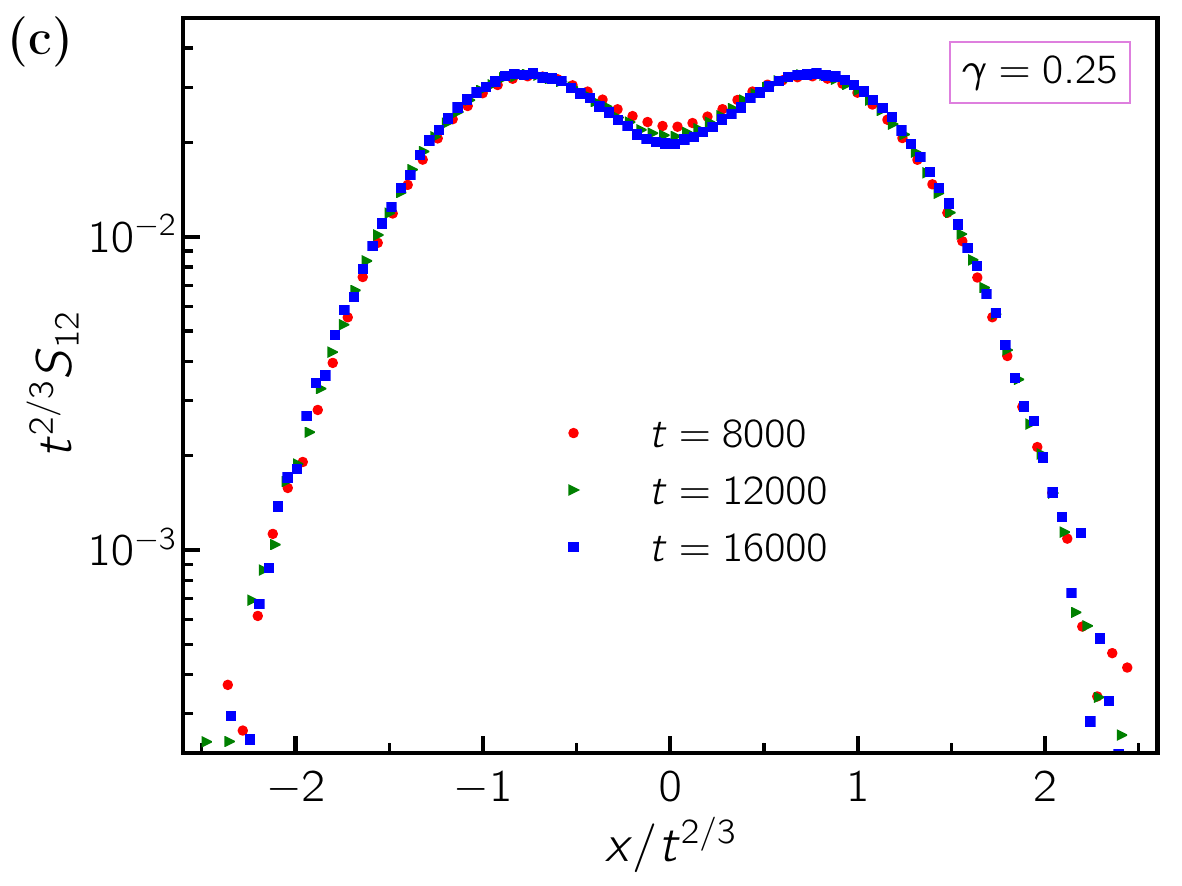}
    \end{subfigure}%
    \begin{subfigure}{0.5\linewidth}
	\includegraphics[width=1\linewidth]{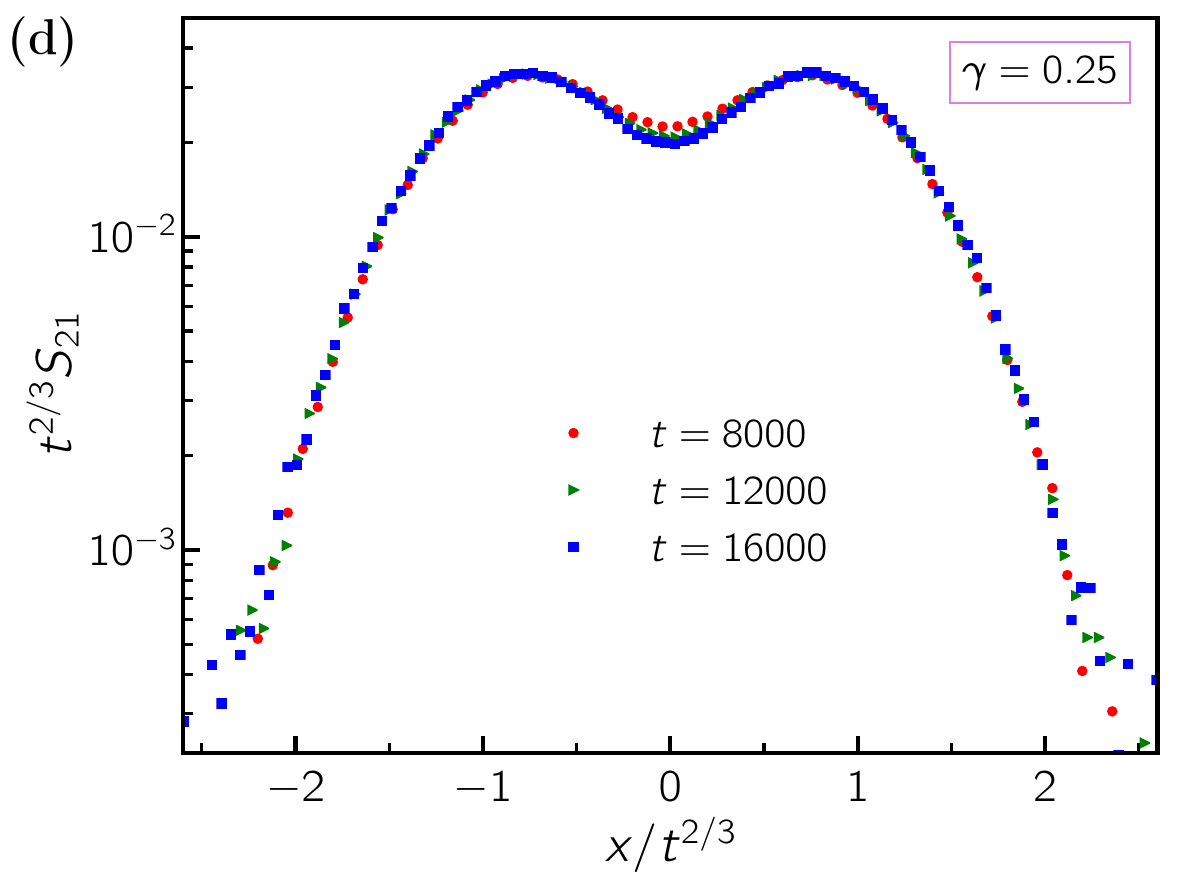}
    \end{subfigure}
    \caption{Plots of the correlator $\mathsfit{S}_{\alpha\beta}$ for the two lanes in the lattice-gas model with $\gamma = 0.25$. The system size is $L=2^{16}$ and the number of independent simulations is $4 \times 10^5$. The initial measure is provided in \equref{4.2} with $\mu = \tilde{\mu} = (\log\gamma)/2$ corresponding to the density $1/2$ in each of the lanes.  In accordance with the sum rule, within a numerical accuracy of $1\%$, $\sum_n \mathsfit{S}_{11} (n,t) = 1 / 4$ and $\sum_n \mathsfit{S}_{12}  (n,t) = 1 /12 $, where $1/4,1/12$
    is the value at $t =0$.  
    }
    \label{fig:4}
\end{figure}

\section{Numerical simulations of the two-lane lattice gas} 
\label{sec5}
\setcounter{equation}{0}

The two-lane model is considered on a ring of $L$ sites, to say $n = 0,\dots, L-1$, with periodic boundary conditions. The initial data are random: at each site independently the distribution \eqref{4.2}  is generated with chemical potentials $\mu = \tilde{\mu} = (\log\gamma)/2$ corresponding to density $1/2$. Rather than a whole range of $\gamma$'s, we choose only $\gamma =0.25$, which corresponds to $\lambda = 0$ and is maximally away from the decoupled $\gamma = 1$. For the dynamics a standard Monte Carlo scheme is used, as described in \cite{2012-popkov-schutz}. System size is $L = 2^{16}$ and simulation time is up to MC $20000$ time steps where a MC time step is $2L$ random sequential updates. Recorded is the 
spacetime correlation matrix
\begin{equation}
    \label{5.1}
    \mathsfit{S}_{\alpha \beta}(n,t) = \big\langle (\eta_{\alpha,n} ( t) - \tfrac{1}{2})( \eta_{\beta,0} (0)- \tfrac{1}{2}) \big\rangle.
\end{equation} 
Numerically  the bracket $\langle \cdot \rangle$ is an average over realizations and lattice points. 
Due to the conservation law the sum rule
\begin{equation}
    \label{5.2}
    \sum_{n=0}^{L-1}\mathsfit{S}(n,t) = C
\end{equation} 
holds.
Exchanging particles and holes is equivalent to exchanging lanes $1$ and $2$. Therefore
\begin{equation}
\label{5.3}
\mathsfit{S}_{11}(n,t) = \mathsfit{S}_{22}(n,t), \quad\mathsfit{S}_{12}(n,t) = \mathsfit{S}_{21}(n,t).
\end{equation} 
Thus only $\mathsfit{S}_{11}$ and $\mathsfit{S}_{12}$ have to be measured. Our results are displayed in \figref{fig:4}, which includes $\mathsfit{S}_{22}$  and $\mathsfit{S}_{21}$ as control check.

The anticipated scaling with the power law $2/3$ is very well confirmed. The sum rule \eqref{5.2} holds with high precision. At $\gamma = 1$, one has two independent TASEPs
and cross correlations vanish. When tuning to $\gamma = 0.25$, $\mathsfit{S}_{12}$ is nonzero and develops a local minimum at $n=0$. The correlator matrix has changed dramatically.
Accidentally, $\mathsfit{S}_{11}$ is rather close to $f_\mathrm{KPZ}$.
These results convincingly establish that the correlator $\mathsfit{S}$ continuously deforms by varying  $\gamma$. There is no evidence that the structure of this $2\times 2$ matrix is related to the single component case. 

\begin{figure}[h]
    \includegraphics[width=1\linewidth]{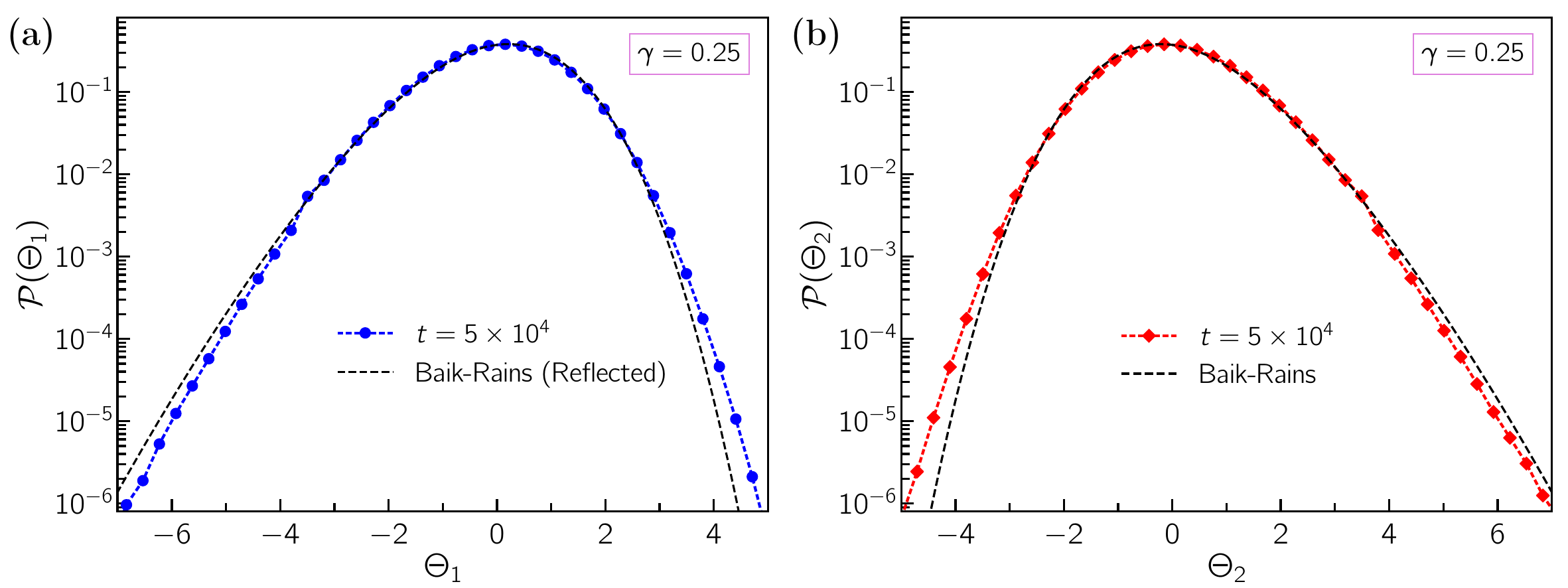}
    \caption{Plots of PDFs for $\Theta_1$ and $\Theta_2$ in the two-lane lattice gas model for $\gamma=0.25$. The initial configurations are chosen from the stationary measure \eqref{4.2} at exact half-filling in each lane. (a) is a mirror image of (b). }
    \label{fig:5}
\end{figure}

 The other observable of interest is the $\alpha$-current, $J_{\alpha,n}(t) $, across the bond $(n,n+1)$ integrated over the time span $[0,t]$. This simply amounts to count the number of exchanges at the bond $(n,n+1)$ in lane $\alpha$ up to time $t$. By particle-hole symmetry, in distribution, 
\begin{equation}
    \label{5.4}
    {J}_{2,n}(t) =  -J_{1,n}(t). 
\end{equation}
Employing  a law of large numbers $J_{\alpha,0}(t) \simeq j_\alpha t$ with 
$j_1 = \gamma^\frac{1}{2}/2(1 + \gamma^\frac{1}{2} )$, $j_2 = -j_1$, 
according to \equref{4.11}. As discussed before, the integrated current fluctuations are expected to be of order $t^\frac{1}{3}$. Hence one studies the limit of large $t$ as
\begin{equation}
    \label{5.5}
    \lim_{t \to \infty} t^{-\frac{1}{3}}(J_{\alpha,0}(t) - j_\alpha t) = \Theta_\alpha,
\end{equation} 
where $\Theta_\alpha$ is still random.

As displayed in \figref{fig:5}, at least numerically a limiting random distribution is established.
For $\gamma = 1$,  the limit $\Theta_1$ is known to be the Baik-Rains distribution.
Our simulation confirms this fact with good precision. On the other hand, for $\gamma = 0.25$,
while the distribution looks qualitatively similar, they are clearly distinct. Quantitatively, for $\gamma = 0.25$  skewness equals $0.278$ and kurtosis $0.237$, which should be compared with the Baik-Rains values of 0.359 and 0.289. This result further supports our scenario of continuous deformation as changing $\gamma$. Particle-hole symmetry implies
$\Theta_1 = -\Theta_2$ in distribution, which is well confirmed.


\section{Universality}
\label{sec6}
\setcounter{equation}{0}

Having numerical data for continuum and discrete models offers the possibility to check for KPZ universality. The parameter correspondence of the two models is given by
\begin{equation}
 \label{6.1}  
 \lambda = 2 - \frac{1}{\sqrt{\gamma}},\qquad b = \frac{1}{2\sqrt{2}}\lambda_2, 
\end{equation}
compare with \eqref{4.20}. For a single component, KPZ scaling theory separates into model-dependent parameters and universal scaling functions. For our case, no such theory is available,
but some aspects can be tested. Following the steps in Section~\ref{sec4}, it is natural to introduce the random field 
\begin{equation}
 \label{6.2}
\varphi_{\alpha,n} = (Q^{-1}R)_{\alpha\beta}(\eta_{\beta,n}- \tfrac{1}{2}). 
\end{equation}
Then, in our context of time stationary initial conditions, KPZ universality means that the statistics of $\varphi_{\alpha,n}(t)$ is close to  
the statistics of the field $\phi_\alpha(x,t)$, as governed by the stochastic Burgers equation \eqref{3.1}. More precisely, long time scaling functions are identical up to model-dependent parameters.

\begin{figure}[h]
    \begin{subfigure}{0.5\linewidth}
	\includegraphics[width=1\linewidth]{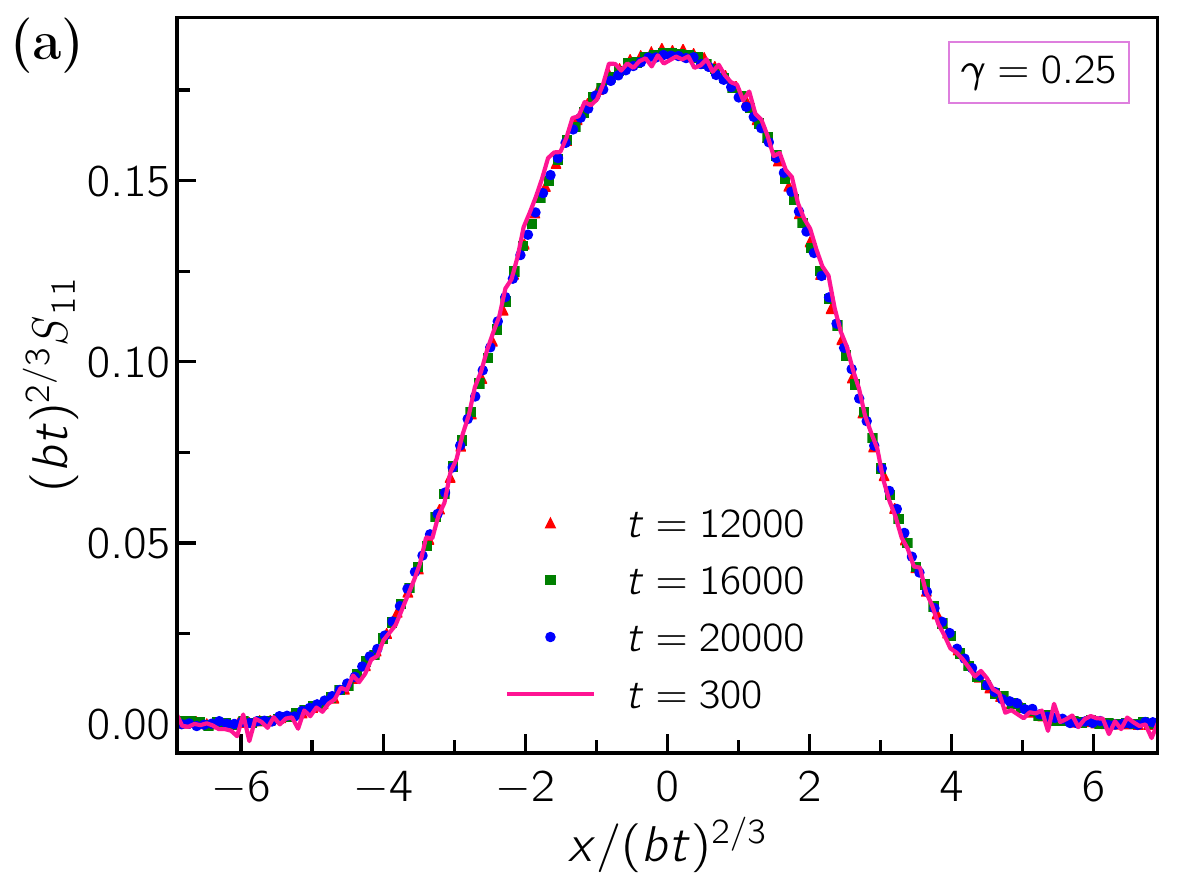}
    \end{subfigure}%
    \begin{subfigure}{0.5\linewidth}
	\includegraphics[width=1\linewidth]{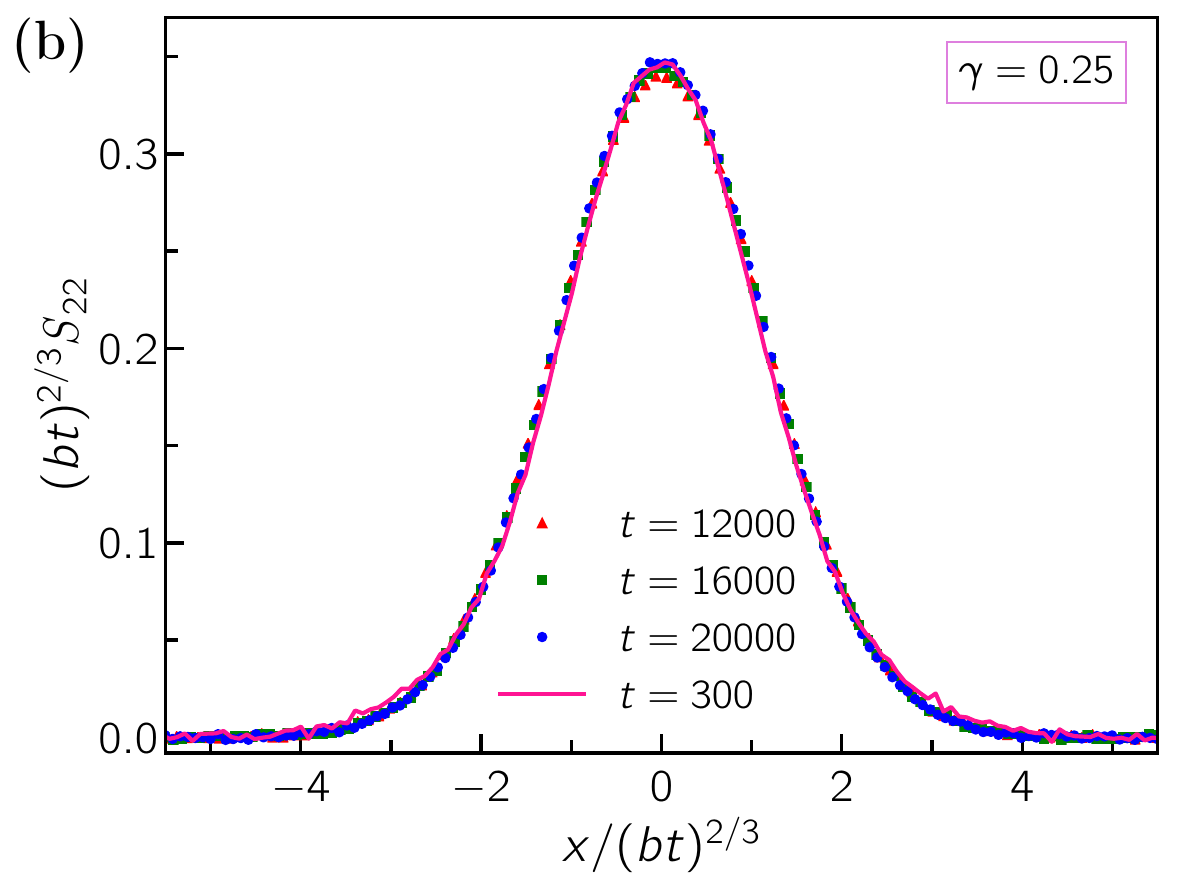}
    \end{subfigure}
    \caption{Comparison of the spacetime correlators (a) $S_{11}$ with
    $P_{11}$ and (b) $S_{22}$ with
    $P_{22}$. 
    The solid lines in deep pink represent the spacetime correlators, $S_{11}$ and $S_{22}$, of the coupled stochastic Burgers equations at $\lambda = 0,\,b=3$. The red, green, and blue filled symbols represent the correlators, $P_{11}$ and $P_{22}$, for the two-lane model at $\gamma = 0.25$. 
    }
    \label{fig:6}
\end{figure}

Our first test concerns the spacetime correlator of  $\varphi_{\alpha,n}(t)$ at $\gamma = 0.25$, which according to Eq. \eqref{5.1} is given by
\begin{equation}
    \label{6.3}
     Q^{-1}R \mathsfit{S}(n,t) R^\mathrm{T}Q^{-1} = P(n,t).
\end{equation}
Its 12 matrix element is very close to zero. The matrix elements $P_{11}(n,t)$, $P_{22}(n,t)$ are now compared
with $S_{11}(x,t)$, $S_{22}(x,t)$, which have been reported already in Section \ref{sec3}. Their asymptotics scales with $b$. Thus we pick $b=3$ as standard and simply replot the two correlators in \figref{fig:6}. For the matrix element $P_{11}(n,t)$, we consider the longest available time and set up an area preserving one-parameter fit defined by 
\begin{equation}
    (a\,t)^{\frac{2}{3}}P_{11}\left((a t)^{2/3} y,t\right), 
\end{equation}
where $y=n/(at)^{2/3}$ is of order $1$. The parameter $a$ is varied to optimize the agreement with the given $S_{11}(x,t)$ and the optimal  $a$ is denoted by $\Lambda_{\|,1}$. In fact it suffices to compare the maxima. Other notions would lead essentially to the same result.   Correspondingly we handle $P_{22}(n,t)$ with a distinct optimal 
parameter $\Lambda_{\|,2}$. As confirmed by \figref{fig:6} this fit works unambiguously. 
 Taking the ``wrong'' value of $\gamma$ or a different linear transformation, the one-parameter fit would fail.
The numerical values at $t=20000$ are $\Lambda_{\|,1} = 0.26$ and $\Lambda_{\|,2} = 0.37 $. 
\begin{center}
\begin{figure}[h]
    \centering
    \includegraphics[width=1\linewidth]{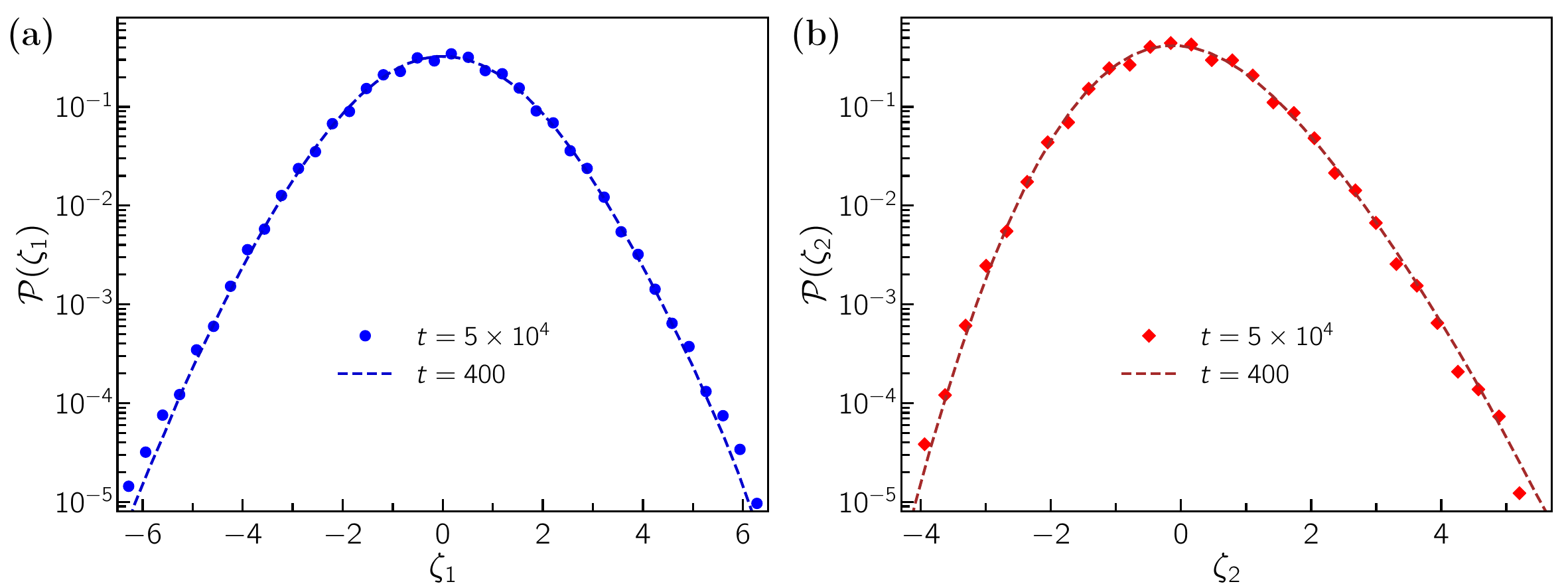}
    \caption{Comparison of the PDFs of fluctuations of the time-integrated current (a) $\zeta_{1}$ with $\Xi_{1}$ and (b) $\zeta_{2}$ with
    $\Xi_2$. The dashed lines represent the PDFs for the coupled stochastic Burgers equations at $\lambda = 0,\, b=4$. The filled symbols represent the data for the two-lane model at $\gamma = 0.25$. 
    }
    \label{fig:7}
\end{figure}
\end{center}

The second test refers to the time-integrated currents as in \eqref{5.5}. According to the rules the transformed random current is
\begin{equation}
    \label{6.4}
    t^{-\frac{1}{3}}(Q^{-1}R)_{\alpha\beta}(J_{\beta,0}(t) - j_\beta t) =\Xi_\alpha(t).
\end{equation}
The free parameter $a$ is inserted in the  PDF at longest time as
\begin{equation}
    \label{6.4}
    a^{-\frac{1}{3}}\,\Xi_\alpha(t), 
\end{equation}
$\alpha = 1,2$. The parameter $a$ is optimized such that the PDF agrees maximally with 
the reference standard, which is the PDF plotted in Fig. \ref{fig:2} with $b=4$. The optimal $a$ is denoted by 
$\Lambda_{\|,\alpha}$. In fact, as before, it suffices to compare the maxima of the PDFs.
The result is displayed in \figref{fig:7}. The numerical values are $\Lambda_{\perp,1} = 2.18$
and $\Lambda_{\perp,2} = 2.74$. 

Giving that only a single parameter is adjusted, for both tests the fit is surprisingly accurate. For our particular choice of parameters, universality is well confirmed. The numerically obtained scaling functions
differ from those known for the one-component system. In this case,  KPZ scaling theory fixes the value of $\Lambda_\|$ and $\Lambda_\perp$.
It would be of interest to better understand their origin in our context.


\section{Mode-coupling theory}
\label{sec7}
\setcounter{equation}{0}

Mode-coupling theory is a self-consistent one-loop approximation for the correlator $S$ of 
\equref{2.14}, see \cite{1985-beijeren--spohn}. With fair success it has been applied to  multi-component KPZ equations
in the non-degenerate setting \cite{2014-spohn}. Thus it is of interest to understand the predictions of this theory for a model with a degenerate flux Jacobian. The derivation of the mode-coupling equations is  discussed in \cite{2014-spohn}. They have to be specialized to a vanishing linear flux term. 

Mode-coupling approximation is usually written in Fourier space for
\begin{equation}
    \label{7.1} 
    \hat{S}(k,t) = \sum_{j \in \mathbb{Z}}\mathrm{e}^{\mathrm{i}kj}S(j,t) .
    \end{equation}
In this representation, the correlator satisfies 
\begin{equation}
    \label{7.2}
    \partial_t \hat{S}(k,t)=  -k^2 \Big(D\hat{S}(k,t) + \int^t_0 ds \hat{M}(k,s) \hat{S}(k,t-s)\Big)
\end{equation}
with memory kernel
\begin{equation}
    \label{7.3}
    \hat{M}_{\alpha\alpha'}(k,s)= 2\int_\mathbb{R}dq\,\mathrm{tr}\big[G^\alpha \hat{S}(k-q,s)^\mathrm{T}G^{\alpha'} \hat{S}(q,s)\big].
\end{equation}
Eq. \eqref{7.2} has to be solved with the initial condition 
\begin{equation}
\label{7.4}
\hat{S}(k,0) =1,
\end{equation}
reflecting the sum rule (2.15) and $C=1$.
One notes that \equref{7.2} is preserved under rotations, i.e. $\tilde{S}= RSR^\mathrm{T}$ satisfies again \equref{7.2} with $\vec{G}$ replaced by $\vec{\tilde{G}}$.

In the long time limit, the solution is expected to be of self-similar form as
\begin{equation} 
    \label{7.5}
    \hat{S}_{\alpha\beta}(k,t)\simeq F_{\alpha\beta}(t^{2/3}k),\quad 
    F_{\alpha\beta}(0) =\delta_{\alpha\beta}.
\end{equation}
Under this scaling the diffusive term \eqref{7.2} is subleading and hence can be omitted. 
However, it might become relevant for special choices of $\vec{G}$. Assuming the approximation \eqref{7.5} and setting  $w = t^{2/3}k$,
the scaling matrix $F$ satisfies
\begin{equation}
    \label{7.6}
    \tfrac{2}{3}w \frac{d}{dw}F(w) = - w^2 \int_0^1 ds\, \mathsf{M}(w,s)F((1-s)^{2/3}w), 
\end{equation}
with initial condition $F(0) = 1$, where 
\begin{equation}
    \label{7.7}
    \mathsf{M}_{\alpha\alpha'}(w,s)
    = 
    2\int_{\mathbb{R}}dq \, \mathrm{tr}\big[G^\alpha F(s^{2/3}(w - q))^\mathrm{T}G^{\alpha'} 
    F(s^{2/3}q)\big].
\end{equation}
 Note that $F$ is even, $F(w) = F(-w)$. 

In principle, $F(w)$ can be obtained by solving \equref{7.2} numerically through iteration, properly including the memory kernel, and taking the long time limit \cite{2013-mendl-spohn}. We have not pursued this direction any further, but focus on the specific  model discussed in Section \ref{sec3}. For this model the off-diagonal matrix elements, $\hat{S}_{12},\hat{S}_{21}$, vanish. 
Such property is correctly reproduced by mode-coupling and Eq. \eqref{7.6} simplifies to two coupled equations for $F_{11}$ and $F_{22}$, which read
\begin{eqnarray}
    \label{7.8}
    &&\hspace{-10pt}   \tfrac{1}{3}w \frac{d}{dw}F_{11}(w) 
    = - 2w^2 \int_0^1 ds \,\mathsf{M}[F_{11},F_{22}](w,s)F_{11}((1-s)^{2/3}w), \\
    &&\hspace{-10pt} \tfrac{1}{3}w \frac{d}{dw}F_{22}(w) 
   = - w^2 \int_0^1 ds \big(\mathsf{M}[F_{11},F_{11}](w,s) 
    +\lambda^2\mathsf{M}[F_{22},F_{22}](w,s)\big)F_{22}((1-s)^{2/3}w),\nonumber
\end{eqnarray}
where $\mathsf{M}[f,\tilde{f}]$ is defined by
\begin{equation}
    \label{7.9}
    \mathsf{M}[f,\tilde{f}](w,s) = 2\int_{\mathbb{R}}dq f(s^{2/3}(w - q))\tilde{f}(s^{2/3}q).    
\end{equation} 
As a shortcoming, the solution depends only on $\lambda^2$, which is not the case for 
the stochastic Burgers equation, compare with the discussion at the end of Section \ref{sec4}.
\begin{center}
\begin{figure}[h!]
    \centering
    \includegraphics[width=0.5\linewidth]{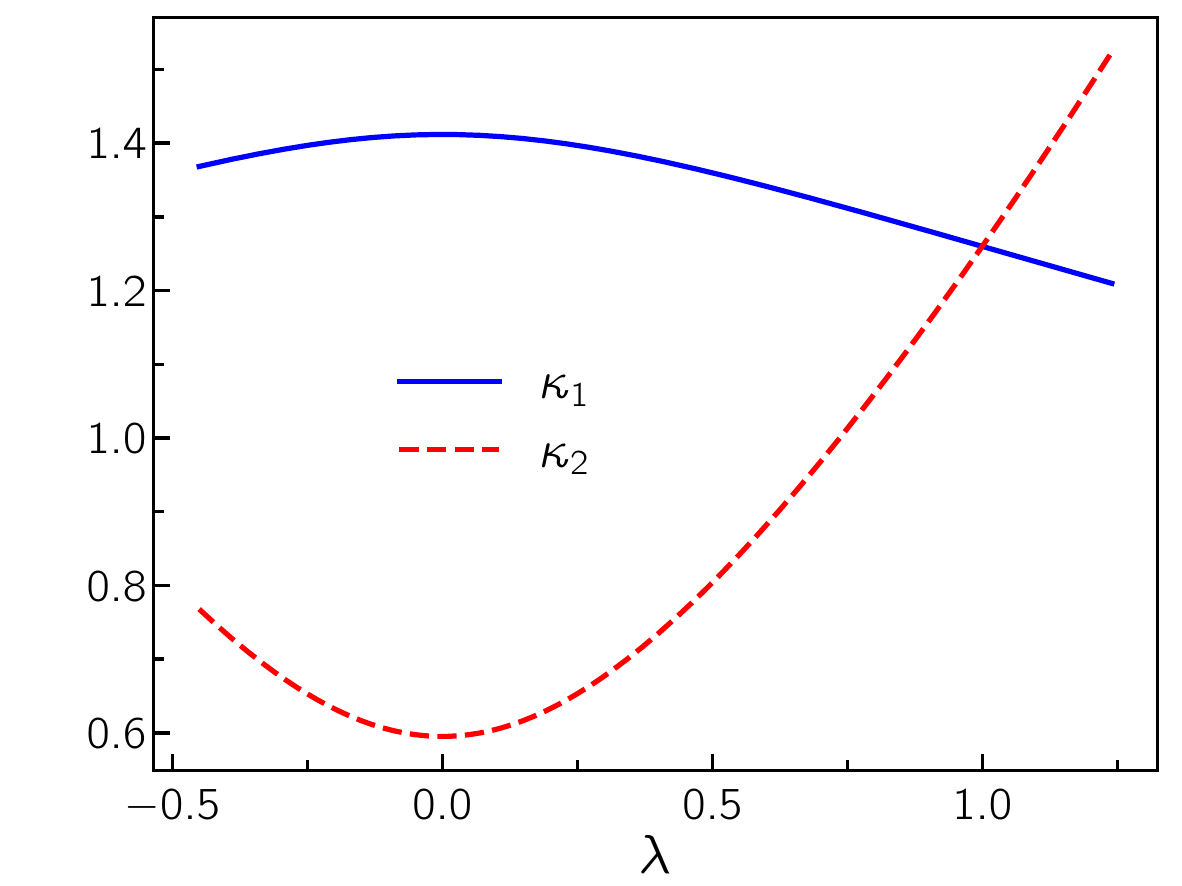}
    \caption{Plots for $\kappa_1$ and $\kappa_2$ versus $\lambda$ obtained by solving \equref{7.11} numerically. At $\lambda=1$, the fixed point is $\kappa_1 = \kappa_2 = 2^{1/3}$. Decreasing $\lambda$,
    the $\kappa$'s are varying with $\kappa_1 < \kappa_2$, in qualitative agreement with the plots in Fig. \ref{fig:1}. The $\kappa$'s are even functions with maximal distance at $\lambda = 0$. }
    \label{fig:8}
\end{figure}
\end{center}
The solution is still complicated and we try the simple-minded ansatz
\begin{equation}
    \label{7.10}
    F_{11}(w) = \exp(-c\kappa_1 w^2 ),\quad F_{22}(w) = \exp(-c\kappa_2 w^2 ),\quad F_{12}(w)=0=F_{21}(w)
\end{equation} 
with some coefficients $c,\kappa_1,\kappa_2>0$. This expression is inserted in Eqs.~\eqref{7.8}, \eqref{7.9}.
An equation for $\kappa_1,\kappa_2$ is obtained by evaluating right and left hand side at $w=0$. Choosing the scale parameter $c$ according to maximal simplicity, the result reads
\begin{equation}
    \label{7.11}
    \kappa_1 =2\frac{1}{\sqrt{\kappa_1 + \kappa_2}}, \quad \kappa_2 = \frac{1}{\sqrt{2\kappa_1}}
    + \lambda^2 \frac{1}{\sqrt{2\kappa_2}}.
\end{equation}
For $\lambda =1$ the solution is $\kappa_1 = \kappa_2$, as it should be and $\kappa_1 = 2^\frac{1}{3}$. As can be seen in \figref{fig:8} as $\lambda$ varies the Gaussian scaling functions vary smoothly. At $\lambda = 0$ the separation of curves is maximal.  Considering the ratio of the maximal heights in $x$ space, one obtains 1.52 on the basis of \eqref{7.11}, while Fig. \ref{fig:1} yields 1.89. A numerical solution of the full mode-coupling equations presumably will do even better .
 
    A slightly more sophisticated argument comes from the observation that at $\lambda = 1$, one has $F_{11} = F_{22} = F_0$, where $F_{0}$ is the fixed point for the one-component mode-coupling equation,
which is numerically known with good precision \cite{2013-mendl-spohn}. One can then expand around this solution.
To first order one obtains non-vanishing corrections to $F_0$. This further supports  that within mode-coupling the scaling functions 
have a nontrivial dependence on $\lambda$.

\section{Discussions}
\label{sec:disc}
\setcounter{equation}{0}
\textit{Relation to the work of Erta\c{s}-Kardar} \cite{1993-ertas-kardar,1992-ertas-kardar}: Using our notation, after transformation to $C=1$, the model considered in \cite{1993-ertas-kardar} has the 
$G^\alpha$ matrices
\begin{equation}
    \label{8.1} 
    G^1 = 
    \begin{pmatrix}
        0 & b\\
        b & 0\\
    \end{pmatrix}, \qquad 
    G^2 = 
    \begin{pmatrix}
        \tilde{b} & 0\\
        0 & d\\
    \end{pmatrix}.
\end{equation}
Cyclicity would require $b=\tilde{b}$, which then corresponds to the line marked by stars in Fig. 2 of \cite{1993-ertas-kardar}. At the time, only scaling exponents were studied. On the other hand, cyclic matrices are only a line in their phase diagram.
Thus it would be of interest to explore also stochastic Burgers equations with non-cyclic coupling matrices.\\\\
\textit{Isotropic spin chains}: 
The $t^\frac{2}{3}$ scaling of the spin-spin correlation has been observed earlier for classical integrable isotropic spin
chains \cite{2019-das--dhar} and in DMRG simulations of a discretized version of the isotropic Heisenberg chain
\cite{2019-ljubotina--prosen}. Furthermore, the scaling function shows excellent agreement with the scaling function $f_\mathrm{KPZ}$ from the single component Burgers equation. 
\equref{3.1} with $b = \tfrac{1}{2}$ was derived in  \cite{2023-nardis--vasseur} as an effective stochastic theory
for spin and helicity density of the spin-spin spacetime correlation function of the XXX spin-$\tfrac{1}{2}$ Heisenberg chain. As in the current contribution, the authors investigate the issue
of universality when varying the parameter $\lambda$. Our numerical simulations indicate that, while the $3/2$ KPZ exponent is confirmed, there are discrepancies from the single component KPZ scaling form. 
Similarly, for time-integrated currents, we observe deviations from the Baik-Rains distribution.  
It is therefore of interest to revisit the issue of an effective stochastic description of isotropic quantum spin chains \cite{2024-krajnik--prosen}.\\\\
\textit{Beta log-gas}: As one of our central findings, the coupled systems under consideration have KPZ exponent but coupling-dependent scaling functions. Although exceptional such features are known from other models in statistical mechanics.
One example is the one-dimensional repulsive log gas with strength parameter
$\beta$. At $\beta = 2$, the Gibbs measure is the joint probability density of the $N$ eigenvalues of a GUE random matrix. The spatial scale is chosen such that the edge of the density of states equals $2N$, i.e. the typical distance of eigenvalues is of order 1. Considering the largest eigenvalue, it is order $N^\frac{1}{3}$  away from the edge. 
For $\beta = 2$, at  large $N$ the limiting distribution is Tracy-Widom \cite{2000-johansson}. Now varying $\beta$, the scaling exponent $\tfrac{1}{3}$ stays put, but the asymptotic probability density for the largest eigenvalue depends on $\beta$ \cite{2011-ramirez--virag, 2010-forrester}.

\section{Summary and Outlook}
\label{sec:sum}
In our contribution, the KPZ scaling theory is extended to stochastic dynamics with two conservation laws possessing a degenerate flux Jacobian. The structure function is then a $2\times 2$ matrix
and its two peaks have the same velocity. The theory is applied to the steady state spacetime correlator and time-integrated current across the origin. Numerically compared is the long time asymptotics 
of a two-lane stochastic lattice gas and
a continuum coupled stochastic Burgers equation.
Our central innovation relies on a similarity transform of the
static susceptibility matrix. 
In our simulations, the dynamical scaling exponent $3/2$
is confirmed with great precision. The form of the scaling functions, however,  
cannot be explained in terms of what one knows about the one-component case.

Many interesting problems remain yet to be explored. We have investigated only a single parameter value. Hence the most immediate issue is a better understanding of scaling functions for two-component systems at the umbilic point. From a theoretical perspective,
for a single component the steady state is expected to have a finite correlation length.
For two components, away from cyclicity, this might no longer be valid. If so, the dynamical exponent 
might change. A further point concerns two-component lattice gases. The two-lane model 
studied here seems to one of the very few examples allowing for degeneracy of the flux Jacobian. Another interesting example is a system comprising of sliding particles on a fluctuating landscape \cite{2017-chakraborty--barma, 2019-chakraborty--barma}. It would be of interest to enlarge the class of such models.

\section*{Acknowledgements} HS greatly benefited from discussions with Jacoppo De Nardis and thanks Fabio Toninelli for mentioning the Beta log-gas. DR, AD and MK acknowledge support from the Department of Atomic Energy, Government of India, under Project No. RTI4001. KK
thanks Alexander von Humboldt Foundation for their support which
made his visit to Munich possible.
KK and MK thank the hospitality of the Department of Mathematics of the Technical University of Munich, Garching (Germany). AD, MK and HS thank the VAJRA faculty scheme (No. VJR/2019/000079) from the Science and Engineering Research Board (SERB), Department of Science and Technology, Government of India.

\appendix
\section{\\Transforming $\vec{G}$-matrices}
\label{app1}
We start from two general matrices $G^1,G^2$ satisfying the cyclicity condition,
\begin{equation}
\label{A.1} 
G^1 = 
\begin{pmatrix}
a & b\\
b & c\\
\end{pmatrix}, \qquad G^2 = 
\begin{pmatrix}
b & c\\
c&d\\
\end{pmatrix}.
\end{equation}
Then under rotations they transform to $\tilde{G}^\alpha$, which explicitly are given by
\begin{eqnarray}
\label{A.2} 
&&\tilde{G}^1_{11} = 
a\cos^3 \vartheta + 3b\cos^2 \vartheta\sin \vartheta +3c \cos \vartheta\sin^2 \vartheta +d \sin^3 \vartheta,\nonumber\\
&& \tilde{G}^1_{12}= \tilde{G}^1_{21}= b\cos^3 \vartheta + (-a+2c)\cos^2 \vartheta\sin \vartheta + (-2b +d)\cos \vartheta\sin^2 \vartheta -c \sin^3 \vartheta,\nonumber\\
&&\tilde{G}^1_{22}= c\cos^3 \vartheta + (-2b +d)\cos^2 \vartheta\sin \vartheta +(a - 2c) \cos \vartheta\sin^2 \vartheta +b \sin^3 \vartheta
\end{eqnarray}
and
\begin{eqnarray}
\label{A.3}
&&\tilde{G}^2_{11} = b\cos^3 \vartheta + (-a + 2c)\cos^2 \vartheta\sin \vartheta +(-2b +d) \cos \vartheta\sin^2 \vartheta -c \sin^3 \vartheta,\nonumber\\
&&\tilde{G}^2_{12}= \tilde{G}^2_{21} = c\cos^3 \vartheta + (-2b +d)\cos^2 \vartheta\sin \vartheta +(a - 2c) \cos \vartheta\sin^2 \vartheta +b \sin^3 \vartheta,\nonumber\\
&&\tilde{G}^2_{22} = d\cos^3 \vartheta -3c\cos^2 \vartheta\sin \vartheta + 3b \cos \vartheta\sin^2 \vartheta - a \sin^3 \vartheta.
\end{eqnarray}
By inspection $\tilde{G}^\alpha$ are still cyclic.
Setting either $b,c,d = 0$ or $a,b,c =0$ yields \eqref{2.29a} and \eqref{2.29b}.

\section{\\Further results for the two-lane model}

\subsection{\\Cyclicity}
\label{app3}
To merely find out about cyclicity, we can also check the one
of $\hat{H}^\alpha$. The inverse of $C$ is given by
\begin{equation}
\label{C.1}
C^{-1} = \frac{1}{2\Omega - 4 \Omega^2}
\begin{pmatrix}
        1 & 1 - 4\Omega\\
        1 - 4\Omega & 1\\
    \end{pmatrix}.
    \end{equation}
Using the definition \eqref{2.12} one obtains
\begin{equation}
\label{C.2}
\hat{H}^1 = (C^{-1})_{11}H^1 + (C^{-1})_{12}H^2,\qquad
\hat{H}^2 = (C^{-1})_{21}H^1 + (C^{-1})_{22}H^2.
\end{equation}
Cyclicity follows by inserting $\vec{H}$ from \eqref{4.9}
and using the identity \eqref{4.10} for $\Omega$.
\subsection{\\Uniqueness of the umbilic point}
\label{app2}
It is convenient to switch to new coordinates
\begin{equation}
\label{B.1}
w = \tfrac{1}{2} - u,\quad  y = \tfrac{1}{2}-v 
\end{equation}
with $|w| < \tfrac{1}{2},|y| < \tfrac{1}{2}$.
Then
 \begin{equation}
 \label{B.2}
     A = 
    \begin{pmatrix}
     2w -y & -w\\
        y & w -2y \\
    \end{pmatrix}
   + 
   \kappa \begin{pmatrix}
        y + \tfrac{1}{2}q(w+y) & w + \tfrac{1}{2}q(w+y)\\
       - y - \tfrac{1}{2}q(w+y) & -w - \tfrac{1}{2}q(w+y)\\
    \end{pmatrix},
\end{equation}
where
\begin{equation}
\label{B.3}
\kappa =  \frac{1}{\sqrt{f }},\quad f(w,y) = 1 +q + q^2(w+y)^2 + 4q wy.
\end{equation}
Proved is the following\\\\
\textbf{Lemma}. If $q > -1 $  and $q \neq 0$, then $u= \tfrac{1}{2} = v$ is the unique umbilic 
point of the flux Jacobian $A$.\\\\
\textit{Proof}: We consider the characteristic polynomial of $A$. Its discriminant is given by
\begin{equation}
\label{B.4}
\mathsfit{D} = (A_{11} - A_{22})^2 + 4 A_{12}A_{21}.   
\end{equation}
One knows already that $\mathsfit{D} \geq 0$. To be established is $\mathsfit{D} = 0$
only if $w= 0 = y$.

One computes
\begin{equation}
\label{B.5}
\mathsfit{D} = (1 +\kappa +q\kappa)^2  (w+y)^2+ 2q\kappa(1 -\kappa -\tfrac{1}{2}q\kappa)(w^2 +y^2)
-4 \big( \tfrac{1}{4}q^2\kappa^2 + (1-\kappa -\tfrac{1}{2}q\kappa)^2\big)wy.
\end{equation}
Clearly, $\mathsfit{D}$ is a quadratic form in $w,y$, which defines the $2\times 2$ matrix $E$.
It remains to show that $E>0$, which is implied by $\mathrm{tr}E  >0$ and $\det E >0$. 
Now
\begin{equation}
\label{B.6}
\mathrm{tr}E = (1+ \kappa)^2 + 4q\kappa, \quad \det E=  16 (1+q)\kappa(1-\kappa)^2.
\end{equation}
For $q > -1$ and $\kappa >0$, one concludes that $\mathrm{tr}E > 0$.  The determinant can vanish only at $\kappa = 1$. Thus we have to discuss the solution of $1 + q(w+y)^2 +4wy = 0$.
Its only solutions are $\tfrac{1}{2}(1,-1),  \tfrac{1}{2}(-1,1)$. In the interior of the quadrant the left hand side is strictly positive and thus $\det E >0$.\\

\section{\\Symmetric and anti-symmetric combinations}
\label{sec:pm}

\begin{center}
\begin{figure}[h]
    \centering
    \begin{subfigure}{0.5\linewidth}
	\includegraphics[width=1\linewidth]{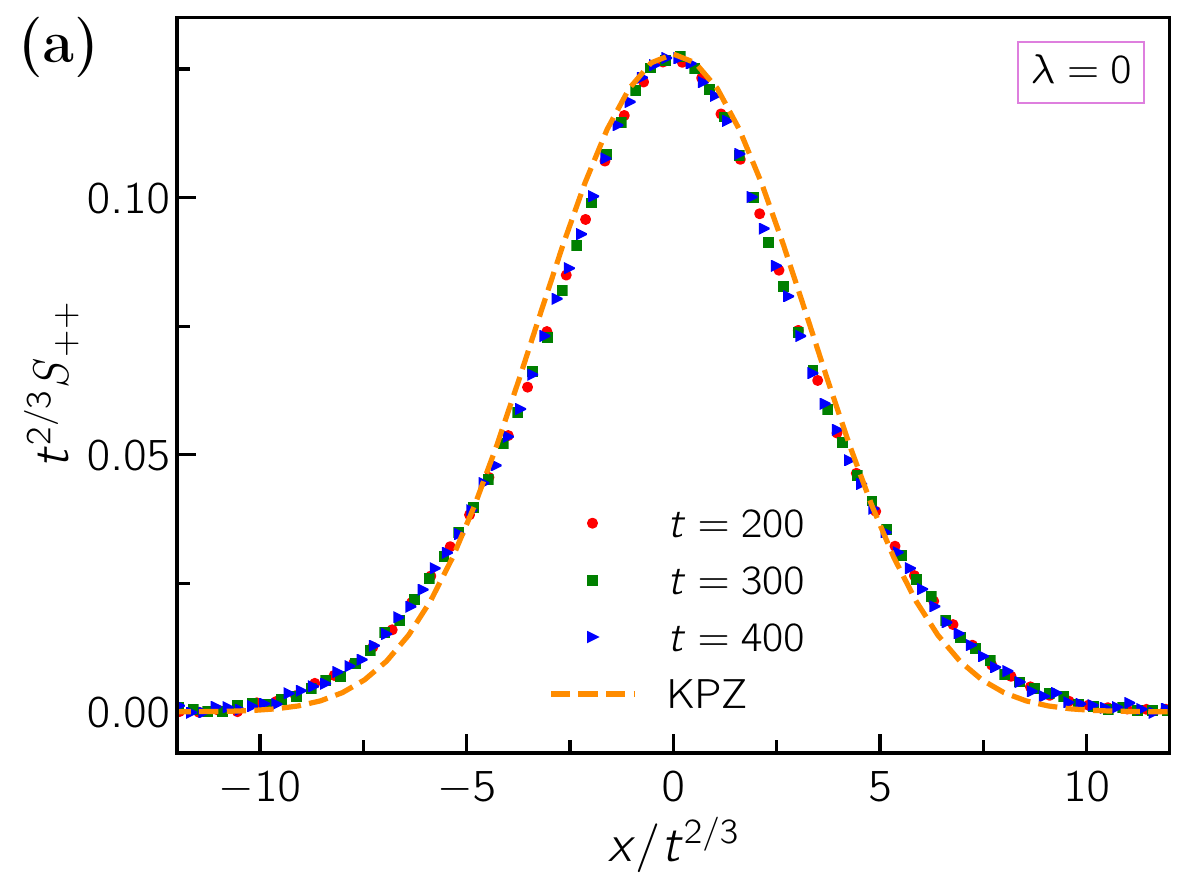}
    \end{subfigure}%
    \begin{subfigure}{0.5\linewidth}
	\includegraphics[width=1\linewidth]{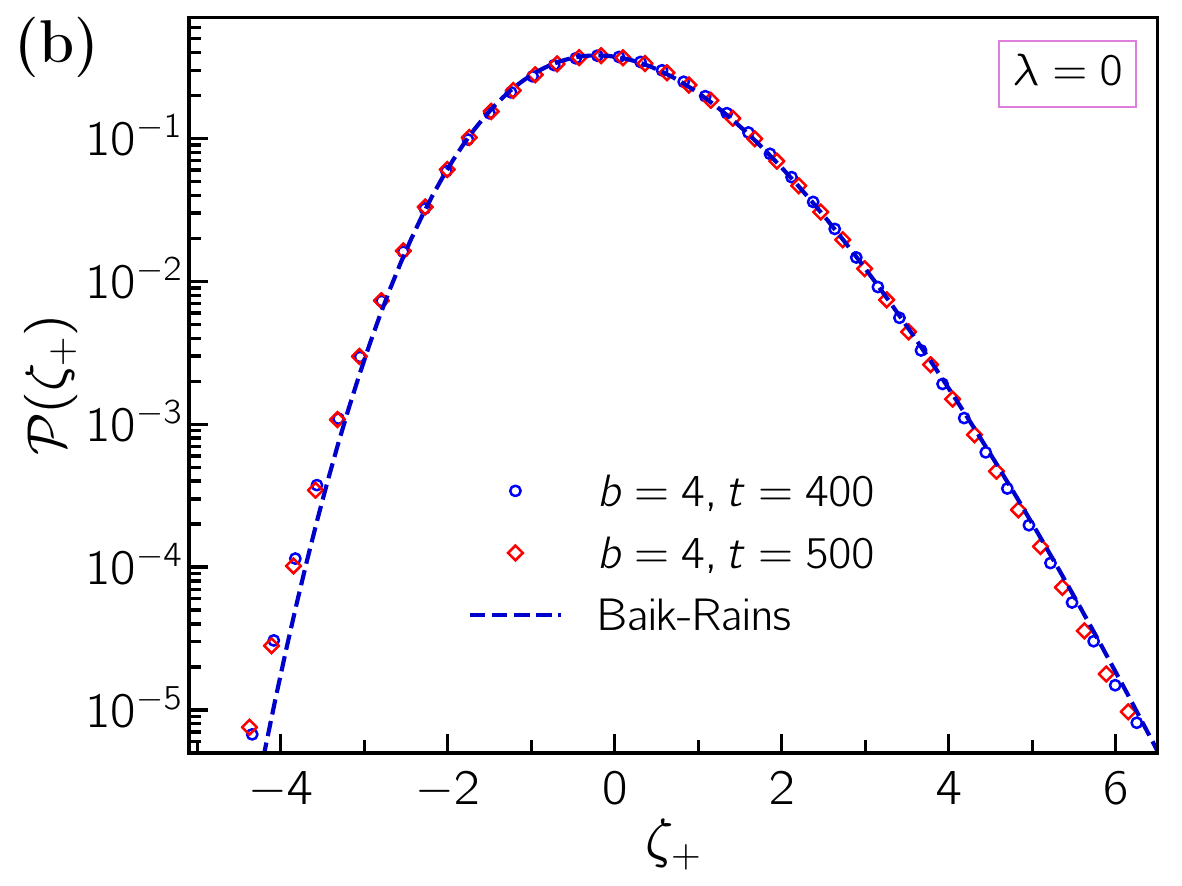}
    \end{subfigure}
    \caption{Plots of (a) the spacetime correlation $S_{++}$ (\ref{app2.14}) and (b) the PDF (\ref{app_zeta_pm}) for the height fluctuations $\zeta_{+}$ corresponding to the field $\phi_+ = (\phi_1 + \phi_2) / \sqrt{2}$. The spacetime correlation for $\phi_{-} = (-\phi_1 + \phi_2) / \sqrt{2}$ is the same as $S_{++}$. Also, the distributions of the height fluctuations corresponding to $\phi_{+}$ and $\phi_{-}$ are same (see Table~\ref{tab:1}) owing to the symmetry $h_1 \to -h_1$. 
    } 
    \label{fig:9}
\end{figure}
\end{center}

If one considers Eq.~(\ref{3.1}) at $\lambda=1$, it is easy to show that symmetric and anti-symmetric combinations, $\phi_{\pm} = (\phi_2\pm \phi_1)/\sqrt{2}$, decouple into two independent stochastic Burgers equations. Hence one may wonder how $\phi_\pm$ changes when $\lambda\neq 1$. In Fig.~\ref{fig:9}, we show spacetime correlations and current fluctuations for the symmetric combination at $\lambda=0$. The spacetime correlation is defined by 
\begin{equation}
    \label{app2.14}
    S_{++}(x,t) = \langle\phi_+(x,t)\phi_+(0,0)\rangle.
\end{equation}
and the current fluctuation are obtained using 
\begin{equation}
 \label{app_zeta_pm}
\zeta_{\pm} = (\zeta_2 \pm \zeta_1)/\sqrt{2}\, . 
\end{equation}

We see that although KPZ exponent is confirmed the scaling functions and current distributions differ significantly from the single component stochastic Burgers equation. 
Therefore, the case $\lambda\neq 1$ is not equivalent to two independent Burgers equations as seems to have been suggested in \cite{2023-nardis--vasseur}.

\section*{References}
\bibliographystyle{unsrt}
\bibliography{2kpz-refs-new.bib}

\end{document}